\RequirePackage[displaymath, mathlines,running]{lineno}
\documentclass[showpacs]{iopart}[floatfix,aps,prd,lengthcheck,superscriptaddress,amssymb,amsmath,amsfonts,nofootinbib,altaffilletter,nopreprintnumbers,showpacs]
\usepackage[latin1]{inputenc}
\usepackage[T1]{fontenc}
\usepackage{acronym}
\usepackage{graphicx}
\usepackage[numbers]{natbib}
\usepackage{multirow}
\usepackage[caption=false]{subfig}
\usepackage{tabularx}
\usepackage[table]{xcolor}
\usepackage{xspace}
\usepackage{pdflscape}
\RequirePackage{doi}
\usepackage{hyperref}
\usepackage[normalem]{ulem}

\newcommand{\inserted}[1]{#1}
\newcommand{\removed}[1]{\iffalse #1 \fi}

\DeclareRobustCommand{\ensuretext}[1]{\ifmmode \textrm{#1} \else #1 \fi}

\definecolor{lightgray}{gray}{0.9} 
\rowcolors{1}{lightgray}{}

\begin{document}

\input{event_macros.tex}
\DeclareRobustCommand{\HONLYTIME}{\ensuremath{10.4}\xspace}

\DeclareRobustCommand{\HTIME}{\ensuremath{244.8}\xspace}

\DeclareRobustCommand{\LONLYTIME}{\ensuremath{9.9}\xspace}

\DeclareRobustCommand{\LTIME}{\ensuremath{251.6}\xspace}

\DeclareRobustCommand{\VONLYTIME}{\ensuremath{25.1}\xspace}

\DeclareRobustCommand{\VTIME}{\ensuremath{248.8}\xspace}

\DeclareRobustCommand{\HONLYPERCENTOFANY}{\ensuremath{3.3\%}\xspace}

\DeclareRobustCommand{\HONLYPERCENTINCZERO}{\ensuremath{3.1\%}\xspace}

\DeclareRobustCommand{\LONLYPERCENTOFANY}{\ensuremath{3.1\%}\xspace}

\DeclareRobustCommand{\LONLYPERCENTINCZERO}{\ensuremath{3.0\%}\xspace}

\DeclareRobustCommand{\VONLYPERCENTOFANY}{\ensuremath{7.9\%}\xspace}

\DeclareRobustCommand{\VONLYPERCENTINCZERO}{\ensuremath{7.5\%}\xspace}

\DeclareRobustCommand{\HLVEXCLUSIVETIME}{\ensuremath{153.0}\xspace}

\DeclareRobustCommand{\HLVEXCLUSIVEPERCENTOFANY}{\ensuremath{48.0\%}\xspace}

\DeclareRobustCommand{\HLVEXCLUSIVEPERCENTINCZERO}{\ensuremath{45.9\%}\xspace}

\DeclareRobustCommand{\HLEXCLUSIVETIME}{\ensuremath{49.7}\xspace}

\DeclareRobustCommand{\HLEXCLUSIVEPERCENTOFANY}{\ensuremath{15.6\%}\xspace}

\DeclareRobustCommand{\HLEXCLUSIVEPERCENTINCZERO}{\ensuremath{14.9\%}\xspace}

\DeclareRobustCommand{\HVEXCLUSIVETIME}{\ensuremath{31.8}\xspace}

\DeclareRobustCommand{\HVEXCLUSIVEPERCENTOFANY}{\ensuremath{10.0\%}\xspace}

\DeclareRobustCommand{\HVEXCLUSIVEPERCENTINCZERO}{\ensuremath{9.5\%}\xspace}

\DeclareRobustCommand{\LVEXCLUSIVETIME}{\ensuremath{39.0}\xspace}

\DeclareRobustCommand{\LVEXCLUSIVEPERCENTOFANY}{\ensuremath{12.2\%}\xspace}

\DeclareRobustCommand{\LVEXCLUSIVEPERCENTINCZERO}{\ensuremath{11.7\%}\xspace}

\DeclareRobustCommand{\HEXCLUSIVETIME}{\ensuremath{10.4}\xspace}

\DeclareRobustCommand{\HEXCLUSIVEPERCENTOFANY}{\ensuremath{3.3\%}\xspace}

\DeclareRobustCommand{\HEXCLUSIVEPERCENTINCZERO}{\ensuremath{3.1\%}\xspace}

\DeclareRobustCommand{\LEXCLUSIVETIME}{\ensuremath{9.9}\xspace}

\DeclareRobustCommand{\LEXCLUSIVEPERCENTOFANY}{\ensuremath{3.1\%}\xspace}

\DeclareRobustCommand{\LEXCLUSIVEPERCENTINCZERO}{\ensuremath{3.0\%}\xspace}

\DeclareRobustCommand{\VEXCLUSIVETIME}{\ensuremath{25.1}\xspace}

\DeclareRobustCommand{\VEXCLUSIVEPERCENTOFANY}{\ensuremath{7.9\%}\xspace}

\DeclareRobustCommand{\VEXCLUSIVEPERCENTINCZERO}{\ensuremath{7.5\%}\xspace}

\DeclareRobustCommand{\HDOUBLETIME}{\ensuremath{81.5}\xspace}

\DeclareRobustCommand{\HDOUBLETIMEPERCENTOFANY}{\ensuremath{25.6\%}\xspace}

\DeclareRobustCommand{\HDOUBLETIMEPERCENTINCZERO}{\ensuremath{24.4\%}\xspace}

\DeclareRobustCommand{\LDOUBLETIME}{\ensuremath{88.7}\xspace}

\DeclareRobustCommand{\LDOUBLETIMEPERCENTOFANY}{\ensuremath{27.8\%}\xspace}

\DeclareRobustCommand{\LDOUBLETIMEPERCENTINCZERO}{\ensuremath{26.6\%}\xspace}

\DeclareRobustCommand{\VDOUBLETIME}{\ensuremath{70.8}\xspace}

\DeclareRobustCommand{\VDOUBLETIMEPERCENTOFANY}{\ensuremath{22.2\%}\xspace}

\DeclareRobustCommand{\VDOUBLETIMEPERCENTINCZERO}{\ensuremath{21.2\%}\xspace}

\DeclareRobustCommand{\ONEDETECTORTIME}{\ensuremath{45.3}\xspace}

\DeclareRobustCommand{\ONEDETECTORPERCENTOFANY}{\ensuremath{14.2\%}\xspace}

\DeclareRobustCommand{\ONEDETECTORPERCENTINCZERO}{\ensuremath{13.6\%}\xspace}

\DeclareRobustCommand{\TWODETECTORTIME}{\ensuremath{120.5}\xspace}

\DeclareRobustCommand{\TWODETECTORPERCENTOFANY}{\ensuremath{37.8\%}\xspace}

\DeclareRobustCommand{\TWODETECTORPERCENTINCZERO}{\ensuremath{36.1\%}\xspace}

\DeclareRobustCommand{\THREEDETECTORTIME}{\ensuremath{153.0}\xspace}

\DeclareRobustCommand{\THREEDETECTORPERCENTOFANY}{\ensuremath{48.0\%}\xspace}

\DeclareRobustCommand{\THREEDETECTORPERCENTINCZERO}{\ensuremath{45.9\%}\xspace}

\DeclareRobustCommand{\ONEORMOREDETECTORTIME}{\ensuremath{318.8}\xspace}

\DeclareRobustCommand{\ONEORMOREDETECTORPERCENTOFANY}{\ensuremath{100.0\%}\xspace}

\DeclareRobustCommand{\ONEORMOREDETECTORPERCENTINCZERO}{\ensuremath{95.6\%}\xspace}

\DeclareRobustCommand{\TWOORMOREDETECTORTIME}{\ensuremath{273.5}\xspace}

\DeclareRobustCommand{\TWOORMOREDETECTORPERCENTOFANY}{\ensuremath{85.8\%}\xspace}

\DeclareRobustCommand{\TWOORMOREDETECTORPERCENTINCZERO}{\ensuremath{82.0\%}\xspace}

\DeclareRobustCommand{\ANALYSISTIMEINCREASE}{\ensuremath{16.6\%}\xspace}

\DeclareRobustCommand{\HOTHREEASENSITIVEDISTANCE}{\ensuremath{108}\,\textrm{Mpc}\xspace}

\DeclareRobustCommand{\LOTHREEASENSITIVEDISTANCE}{\ensuremath{135}\,\textrm{Mpc}\xspace}

\DeclareRobustCommand{\VOTHREEASENSITIVEDISTANCE}{\ensuremath{45}\,\textrm{Mpc}\xspace}

\DeclareRobustCommand{\HOTHREEBSENSITIVEDISTANCE}{\ensuremath{115}\,\textrm{Mpc}\xspace}

\DeclareRobustCommand{\LOTHREEBSENSITIVEDISTANCE}{\ensuremath{133}\,\textrm{Mpc}\xspace}

\DeclareRobustCommand{\VOTHREEBSENSITIVEDISTANCE}{\ensuremath{51}\,\textrm{Mpc}\xspace}

\DeclareRobustCommand{\HSNGLVTINCREASEESTIMATE}{\ensuremath{1.5\%}\xspace}

\DeclareRobustCommand{\LSNGLVTINCREASEESTIMATE}{\ensuremath{2.5\%}\xspace}

\DeclareRobustCommand{\VSNGLVTINCREASEESTIMATE}{\ensuremath{0.3\%}\xspace}

\DeclareRobustCommand{\SNGLVTINCREASEESTIMATE}{\ensuremath{4.2\%}\xspace}

\DeclareRobustCommand{\HVCOINCVTINCREASEESTIMATE}{\ensuremath{3.3\%}\xspace}

\DeclareRobustCommand{\LVCOINCVTINCREASEESTIMATE}{\ensuremath{8.3\%}\xspace}

\DeclareRobustCommand{\TOTALCOINCVTINCREASEESTIMATE}{\ensuremath{11.6\%}\xspace}

\DeclareRobustCommand{\TOTALVTINCREASEESTIMATE}{\ensuremath{15.8\%}\xspace}

\DeclareRobustCommand{\VTBINONELIMITS}{\ensuremath{1.30 \Msun{} < \mchirp{} \leq 2.70 \Msun{}}\xspace}

\DeclareRobustCommand{\VTBINTWOLIMITS}{\ensuremath{2.70 \Msun{} < \mchirp{} \leq 4.35 \Msun{}}\xspace}

\DeclareRobustCommand{\VTBINTHREELIMITS}{\ensuremath{4.35 \Msun{} < \mchirp{} \leq 8.00 \Msun{}}\xspace}

\DeclareRobustCommand{\VTBINFOURLIMITS}{\ensuremath{8.00 \Msun{} < \mchirp{} \leq 32.00 \Msun{}}\xspace}

\DeclareRobustCommand{\VTBINFIVELIMITS}{\ensuremath{32.00 \Msun{} < \mchirp{} \leq 64.00 \Msun{}}\xspace}

\DeclareRobustCommand{\VTBINSIXLIMITS}{\ensuremath{64.00 \Msun{} < \mchirp{} \leq 128.00 \Msun{}}\xspace}

\DeclareRobustCommand{\VTRATIOBROADALL}{\ensuremath{1.11 \pm 0.01}\xspace}

\DeclareRobustCommand{\VTRATIOBROADBINONE}{\ensuremath{1.14 \pm 0.01}\xspace}

\DeclareRobustCommand{\VTRATIOBROADBINTWO}{\ensuremath{1.14 \pm 0.01}\xspace}

\DeclareRobustCommand{\VTRATIOBROADBINTHREE}{\ensuremath{1.13 \pm 0.02}\xspace}

\DeclareRobustCommand{\VTRATIOBROADBINFOUR}{\ensuremath{1.13 \pm 0.01}\xspace}

\DeclareRobustCommand{\VTRATIOBROADBINFIVE}{\ensuremath{1.08 \pm 0.01}\xspace}

\DeclareRobustCommand{\VTRATIOBROADBINSIX}{\ensuremath{1.03 \pm 0.01}\xspace}

\DeclareRobustCommand{\VTRATIOBBHALL}{\ensuremath{1.16 \pm 0.01}\xspace}

\DeclareRobustCommand{\VTRATIOBBHBINONE}{\ensuremath{--}\xspace}

\DeclareRobustCommand{\VTRATIOBBHBINTWO}{\ensuremath{--}\xspace}

\DeclareRobustCommand{\VTRATIOBBHBINTHREE}{\ensuremath{1.20 \pm 0.02}\xspace}

\DeclareRobustCommand{\VTRATIOBBHBINFOUR}{\ensuremath{1.19 \pm 0.01}\xspace}

\DeclareRobustCommand{\VTRATIOBBHBINFIVE}{\ensuremath{1.15 \pm 0.01}\xspace}

\DeclareRobustCommand{\VTRATIOBBHBINSIX}{\ensuremath{1.12 \pm 0.01}\xspace}

\title[]{Establishing significance of gravitational-wave signals from a single observatory in the PyCBC offline search}

\author{Gareth S. Cabourn Davies}
\address{University of Portsmouth, Portsmouth, PO1 3FX, United Kingdom}
\ead{gareth.davies@port.ac.uk}

\author{Ian W. Harry}
\address{University of Portsmouth, Portsmouth, PO1 3FX, United Kingdom}



\begin{abstract}
Gravitational-wave observations of compact binary coalescences are allowing us to see black holes and neutron stars further into the universe and recent results represent the most sensitive searches for compact objects ever undertaken.
Most searches for gravitational waves from compact binary coalescence currently rely on detecting coincident triggers from multiple detectors.
In this paper, we describe a new method for extrapolating significance of single-detector signals beyond the live-time of the analysis.
Using this method, we can recover loud signals which only triggered in a single detector.
We demonstrate this method in a search of O3 data, and recover $\NUMSNGLEVENTS$ single-detector events with a false alarm rate less than two per year.
These were the same events as discovered in the GWTC-2.1 and GWTC-3 searches in a single detector, and all but one event from 3-OGC and 4-OGC.
Through a campaign of injected signals, we estimate that the total time--volume sensitivity increases by a factor of up to \VTRATIOBBHBINTHREE at a false alarm rate of one per two years compared to completely ignoring single-detector events.

\end{abstract}

\maketitle

\acrodef{FAR}[FAR]{false alarm rate}
\acrodef{GW}[GW]{gravitational wave}
\acrodef{O4}[O4]{the fourth observing run}
\acrodef{O3}[O3]{the third observing run}
\acrodef{O3a}[O3a]{the first part of \ac{O3}}
\acrodef{O3b}[O3b]{the second part of \ac{O3}}
\acrodef{O2}[O2]{second observing run}
\acrodef{O1}[O1]{first observing run}
\newcommand{\PYCBCBROAD}{PyCBC-broad\xspace}
\newcommand{\PYCBCBBH}{PyCBC-BBH\xspace}
\acrodef{CBC}[CBC]{compact binary coalescence}

\acrodef{BNS}[BNS]{binary neutron star}
\acrodef{BBH}[BBH]{binary black hole}
\acrodef{NSBH}[NSBH]{neutron star-black hole binary}

\newcommand{\PYCBC}{PyCBC\xspace}
\newcommand{\GSTLAL}{GstLAL\xspace}

\acrodef{FAR}[FAR]{false alarm rate}
\acrodef{SNR}[SNR]{signal-to-noise ratio}
\newcommand{\VT}{\ensuremath{\langle VT \rangle}}

\newcommand{\p}{\ensuremath{\textrm{p}}}

\newcommand{\newsnr}{\hat{\rho}}
\newcommand{\Msun}{M_\odot}
\newcommand{\pastro}{\ensuremath{\p_\textrm{astro}}}
\newcommand{\mchirp}{\ensuremath{\mathcal{M}}}

\acrodef{GWTC}[GWTC]{gravitational-wave transient catalog}
\acrodef{3OGC}[3-OGC]{the third open gravitational-wave catalog}
\acrodef{LSC}[LSC]{LIGO Scientific Collaboration}
\acrodef{LVC}[LVC]{LIGO Scientific and Virgo Collaboration}
\acrodef{LVK}[LVK]{LIGO Scientific, Virgo and KAGRA Scientific}
\acrodef{aLIGO}{Advanced Laser Interferometer Gravitational-Wave Observatory}
\acrodef{aVirgo}{Advanced Virgo}
\acrodef{LIGO}[LIGO]{Laser Interferometer Gravitational-Wave Observatory}

\acrodef{PSD}[PSD]{power spectral density}

\acrodef{HLV}[HLV]{Hanford--Livingston--Virgo}
\acrodef{HL}[HL]{Hanford--Livingston}
\acrodef{HV}[HV]{Hanford--Virgo}
\acrodef{LV}[LV]{Livingston--Virgo}
\acrodef{H}[H]{LIGO-Hanford}
\acrodef{L}[L]{LIGO-Livingston}
\acrodef{V}[V]{Virgo}

\acrodef{NSF}[NSF]{National Science Foundation}
\acrodef{STFC}[STFC]{Science and Technology Funding Council}

\section{Introduction}
\label{sec:intro}
Gravitational-wave searches and observations of compact binary coalescences are helping us to understand more and more about the universe.
Signals detected in the Advanced \ac{LIGO}~\citep{TheLIGOScientific:2014jea} and Advanced Virgo~\citep{TheVirgo:2014hva} have allowed us to see more black holes and neutron stars, in more configurations and at further distances than previously \inserted{observed}\removed{allowed}.

In order to detect and understand these objects, we must first find them within the data.
This is done in two ways; in low-latency (e.g.~\citep{DalCanton:2020vpm}), which aims to rapidly detect signals which can be disseminated to the wider astronomical community to enable follow-up observations, as seen for GW170817~\citep{LIGOScientific:2017ync}.
Additional analyses are performed offline (e.g.~\citep{LIGOScientific:2020ibl,LIGOScientific:2021djp,Nitz:2021uxj,Nitz:2021zwj}), which aim to more accurately assess the significance of the candidate events, in order to gain a more confident list of events for use in further study such as parameter estimation~\citep{Biwer:2018osg,Berry:2014jja,Cutler:1994ys}, population analyses~\citep{Roulet:2020wyq,LIGOScientific:2018jsj,LIGOScientific:2021psn} and tests of general relativity~\citep{LIGOScientific:2021sio}.

Multiple search analyses are used for \ac{GW} searches, using different search methods and configurations in order to ensure that we remain able to detect any possible \acp{CBC} in the data.
The most sensitive searches for \acp{GW} from \acp{CBC}, both in low-latency and in offline searches~\citep{Davies:2020tsx,Nitz:2017svb,Usman:2015kfa,Babak:2012zx,Messick:2016aqy,Sachdev:2019vvd,Aubin:2020goo,Adams:2015ulm,Hooper:2011rb,Venumadhav:2019lyq}, are based on comparing the data to a bank of waveform templates ~\citep{Roy:2017qgg,Roy:2017oul}.
Additionally, unmodelled searches for coherent excess power between detectors~\citep{LIGOScientific:2016fbo,Klimenko:2015ypf} can find many of the \ac{CBC} signals, as well as loud transient signals not from \acp{CBC}.

\PYCBC is a suite of mainly python-based software for use in the analysis of gravitational-wave data, consisting of a highly modular and configurable set of libararies for searches and parameter estimation of \acp{CBC}~\citep{alex_nitz_2021_4849433}.
\PYCBC workflows take advantage of diverse computational resources including local clusters, XSEDE, and the Open Science Grid~\citep{Weitzel:2017ocs} using the Pegasus workflow management system~\citep{deelman-fgcs-2015}.
\PYCBC software includes low-latency and offline \ac{CBC} searches~\citep{DalCanton:2020vpm,Davies:2020tsx,Nitz:2017svb,Usman:2015kfa}, and parameter estimation \cite{Biwer:2018osg}.
This paper utilises the \PYCBC offline search for \acp{GW} from \acp{CBC}.

\begin{figure*}
\begin{center}
\includegraphics[width=0.6\textwidth]{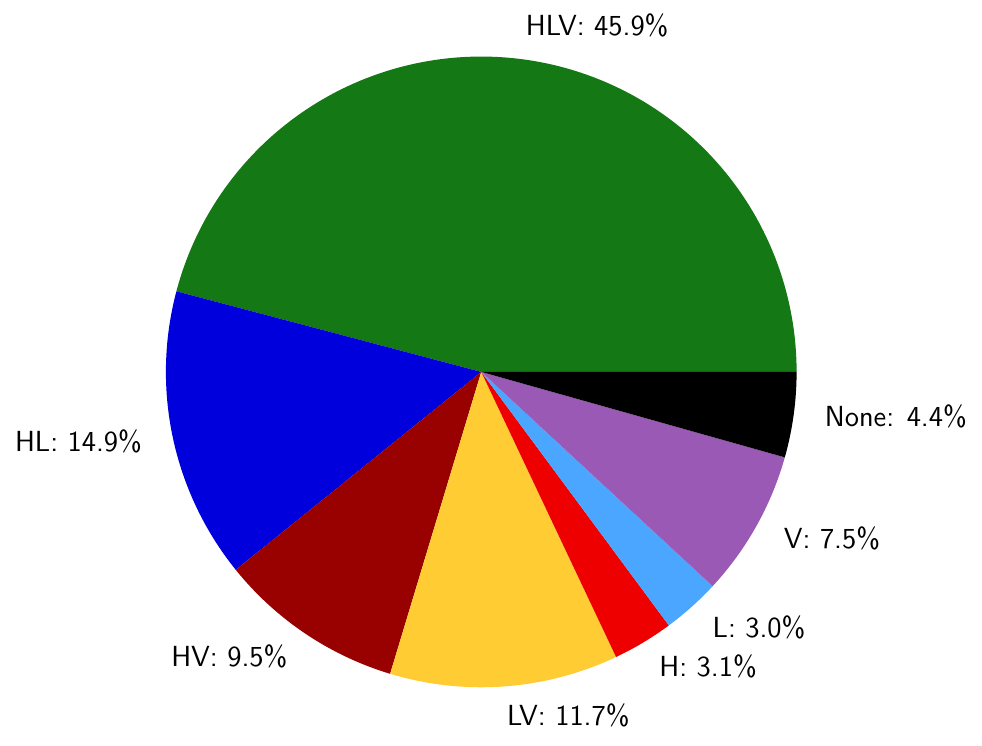}
\end{center}
\caption{\label{fig:science_time}
The fraction of time during O3 for which each combination of detectors was observing, with each detector combination signified by its initials, LIGO-Hanford (H), LIGO-Livingston (L) and Virgo (V).
Times do not include the month-long commissioning break in October 2019.
We see that a significant fraction of the observing run time is when a single observatory is operating (\ONEDETECTORPERCENTINCZERO), or where one of the LIGO observatories is coincident with Virgo only (\VDOUBLETIMEPERCENTINCZERO).
}
\end{figure*}

\inserted{Gravitational-wave detectors do not have a one hundred percent duty cycle, and so we must consider detection of events which occur outside of times when all detectors are operating as designed.}
Figure~\ref{fig:science_time} shows \removed{the times when each of the LIGO detectors and Virgo was operating during \ac{O3}, and} the fraction of time \inserted{that}\removed{with the} different combinations of \removed{active }detectors\inserted{ at LIGO-Hanford, LIGO-Livingston and Virgo were active during \ac{O3}}.

\removed{Both modelled and unmodelled searches generally require data from two or more detectors and generate possible events based on coincident triggers.}
\inserted{Modelled searches generally use peaks in \ac{SNR} time-series, \emph{triggers}, from each detector and coincident sets of these triggers are matched with one another to form \emph{events}, which are candidate signals.}
Standard offline \PYCBC searches for \acp{GW}, for example, generate coincidences from multiple detectors and then the \ac{FAR} is estimated by counting the number of higher-ranked events in a background manufactured from time shifts~\citep{Davies:2020tsx,Nitz:2017svb,Usman:2015kfa}.
There are however a few notable exceptions to this requirement for coincidence in \ac{GW} searches.

An alternative search pipeline, \GSTLAL, also has offline and low latency searches.
The \GSTLAL searches assign significance to single-detector events based on likelihood estimates calculated through distributions of triggers in the individual detectors~\citep{Sachdev:2019vvd}.
A `singles penalty' is applied to this likelhood in order to penalise a signal that is not seen in multiple detectors, \citep{LIGOScientific:2021usb}.

Low latency \PYCBC searches have also recently introduced a method to estimate \acp{FAR}~\citep{DalCanton:2020vpm}, based on extrapolating the background using a fit to an exponential decrease in the number of triggers at a given single-detector ranking statistic.
This method is designed to provide alerts in low latency for single-detector events that could have multimessenger counterparts, and so insists on strict event selection cuts which exclude most \ac{BBH} mergers.

Recent work means that the \PYCBC offline search is able to provide estimates of probability of astrophysical origin \pastro{} for single-detector events~\citep{Nitz:2020naa}.
However \pastro{} should be considered as part of the wider context of all figures of merit of the significance of the candidate event.
For example, \pastro{} can have significant uncertainty when the underlying signal rate is unknown \citep{Andres:2021vew}, and particularly for previously undetected populations of \ac{CBC} (e.g. events with extreme mass ratio).
As a result, we wish to also obtain an estimate of the \ac{FAR} in order to provide further information on the event.

In this paper, we will discuss events in different combinations of detectors in the gravitational wave detector network.
Detectors will be referred to by indicative letters, \ac{H}, \ac{L} and \ac{V}.
The network of detectors will be referred to by the combined initials, \ac{HLV}, \ac{HL}, \ac{HV} and \ac{LV}.

The discussion requires two notions of the detector network, the first of which is the network of detectors which is active at the time of the event; we will refer to this as an event being in the time of the detectors, e.g. \emph{HL time} refers to a time when LIGO-Hanford and LIGO-Livingston are operating, but not Virgo.
When discussing the time in which exactly one detector is operating, this is \emph{single-detector time}, similarly \emph{double-} and \emph{triple times} refer to when exactly two or three detectors are operating.
Times where one-or-more detectors are operating is \emph{any-detector time}, and where two or more detectors is operating is \emph{coincident time}.

The other network we will discuss is the network of detectors which \removed{triggered and}\inserted{generated triggers that} contributed to the event; this is a subset of the active network, and contains only the detectors which contributed to the significance of the event.
We will refer to events of this nature as being, e.g. \emph{HV events} from LIGO-Hanford and Virgo, where these detectors' triggers contribute to the event.
Again, these events can be referred to as \emph{single-detector events}, \emph{double} or \emph{triple} events, and \emph{coincident} events, with equivalent definitions to the same terms for the active network.

By adding the ability to detect single-detector events, we gain sensitivity in a number of situations:
The first is in the single-detector time, which was the case for \ONEDETECTORTIME days, or \ONEDETECTORPERCENTINCZERO of \ac{O3}.
By allowing the use of single-detector time, in \ac{O3} we increase the time available to the search from \TWOORMOREDETECTORTIME to \ONEORMOREDETECTORTIME days, an increase of \ANALYSISTIMEINCREASE.
This will not directly correspond to an increase of the same amount in the sensitive volume--time of the search, as the additional time would need to be weighted by the sensitive distance of the available network.

Secondly, we gain sensitivity where a signal is only observed in a single detector but other detectors are operating and did not see the signal.
For example, the detection of \FULLNAME{GW200105} during \ac{O3b}, where the signal was not seen in Virgo but was in \ac{LV} time.
This was because the signal was not strong enough to be seen in that detector.
In O3, we were therefore more likely to have single-detector detections during two-detector time when one of the operating detectors was Virgo, as Virgo was less sensitive than the \ac{LIGO} detectors.
As the two LIGO detectors had similar sensitivities to one another, it is unlikely that a signal would be loud enough to be seen in one detector but not the other.
In Figure~\ref{fig:science_time} we see that the detector network had one of the \ac{LIGO} detectors coincident with only Virgo for \VDOUBLETIME days, or \VDOUBLETIMEPERCENTOFANY of the any-detector time.


Where we have significantly mismatched sensitivities of detectors in a coincident search, we are presented with an additional problem.
When calculating significance, we remove confident detections from the estimated background, however we cannot do so for these single-detector events using current methods.
If the signal cannot be removed from the background, it can cause contamination in the background used for significance estimates, and cause false alarm rates to be overestimated.

We present here a method for estimating false alarm rates for use in the \PYCBC offline search, explaining how we extrapolate beyond the noise background limit in Section~\ref{sec:extrapolation}, and how this significance is used within a wider search analysis in Section~\ref{sec:combining-far}.
In Section~\ref{sec:o3_results} we present results of a search on \ac{O3} data using this method, comparing to a coincident-only search similar to the \PYCBC \ac{GWTC} analyses \ac{GWTC}-2.1 for \ac{O3a} and \ac{GWTC}-3 for \ac{O3b}.
Section~\ref{sec:sensitivity} then describes the results of associated injection campaigns for estimates of the increase in sensitivity given by the search.

\inserted{The results presented here are for the \PYCBCBROAD and \PYCBCBBH searches as presented in \ac{GWTC}-2.1 and \ac{GWTC}-3; these are searches for a broad range of compact binary coalescence parameters, and for systems which are similar to the majority of previously detected \ac{BBH} systems respectively, with details given in~\citep{LIGOScientific:2020ibl,LIGOScientific:2021djp}.}

\section{Going beyond the edge of the noise distribution}
\label{sec:extrapolation}
In order to assign a false alarm rate to a candidate event, we compare a ranking statistic - a measure of how signal-like we think the event is - to a background, counting the rate of background events ranked higher than our candidate event.
\PYCBC coincident searches use time shifts, where the triggers from each detector are shifted relative to the others in order to build up a background which is entirely made up of events which cannot possibly be real, as the time shifts are far greater than the time taken for the \ac{GW} to travel between detectors.
However when considering the background of single-detector events, we cannot build up this background through time shifts, and so we need an alternative method to \removed{generate}\inserted{approximate} the background.

Without any form of extrapolation, the highest-ranked single-detector event in a search would have a \ac{FAR} of one per the live time of the search under the assumption that it, and all lower-ranked \removed{triggers}\inserted{events}, are noise.
The \GSTLAL-based \ac{CBC} search uses extrapolation based on a KDE approach \citep{Messick:2016aqy}.
Other methods use signal and noise populations to estimate a \pastro{} based on assumptions of the noise distribution beyond the loudest-ranked \removed{trigger}\inserted{event}.
One option for this is to use the expected signal distribution, normalised to assume a certain number of signals above the second-loudest noise candidate \citep{Callister:2017urp}.
Alternatively, one can form the noise distribution by assuming it is proportional to the signal density function at high statistic, normalised so that one noise event is expected between the event and next-loudest event \citep{Nitz:2020naa}.

Here we outline a method to estimate single-detector event significance beyond this limit without assuming any signal distribution through simple extrapolation of the background.
As we do not assume any signal distribution, we are able to estimate the \ac{FAR} based solely on the noise distribution, which is more appropriate where the signal distribution is unknown or poorly-known.
\inserted{A single-detector event cannot a priori be considered as signal or noise, and so we estimate the bulk distribution of events and assume that it is dominated by noise events; this is a safe assuption in current detectors.}

Searches will assign a low \ac{FAR} to \removed{a trigger}\inserted{an event} which is rare compared to other \removed{triggers}\inserted{events} in the analysis.
The most common \removed{triggers}\inserted{events} in the data \removed{are}\inserted{come from} Gaussian noise triggers, which have well-described statistical properties.
Signals in the data occur rarely in the analysis, and have different properties to the Gaussian\removed{ triggers}\inserted{-produced events}, meaning that they stand out from the background.
However, non-Gaussian transient glitches in the data can also produce triggers which stand out from the background, some of which mimic certain types of \ac{CBC} signal \citep{Cabero:2019orq}.

Most glitches are removed by coincidence requirements; as the glitch times are uncorrelated between detectors they will be removed from consideration as possible events.
Without this coincidence requirement, however, non-Gaussian transients could be found to be very significant.

\subsection{Removal of confident noise events}

As a result of these additional difficulties in extrapolating single-detector significance, we will place more stringent limits on certain properties of the candidate events\inserted{ and their associated triggers}.

Firstly, in order to prevent unneccesary computational effort, we make a cut on triggers which we can be confident are a part of the noise.

\begin{figure}
\begin{center}
\includegraphics[width=\textwidth]{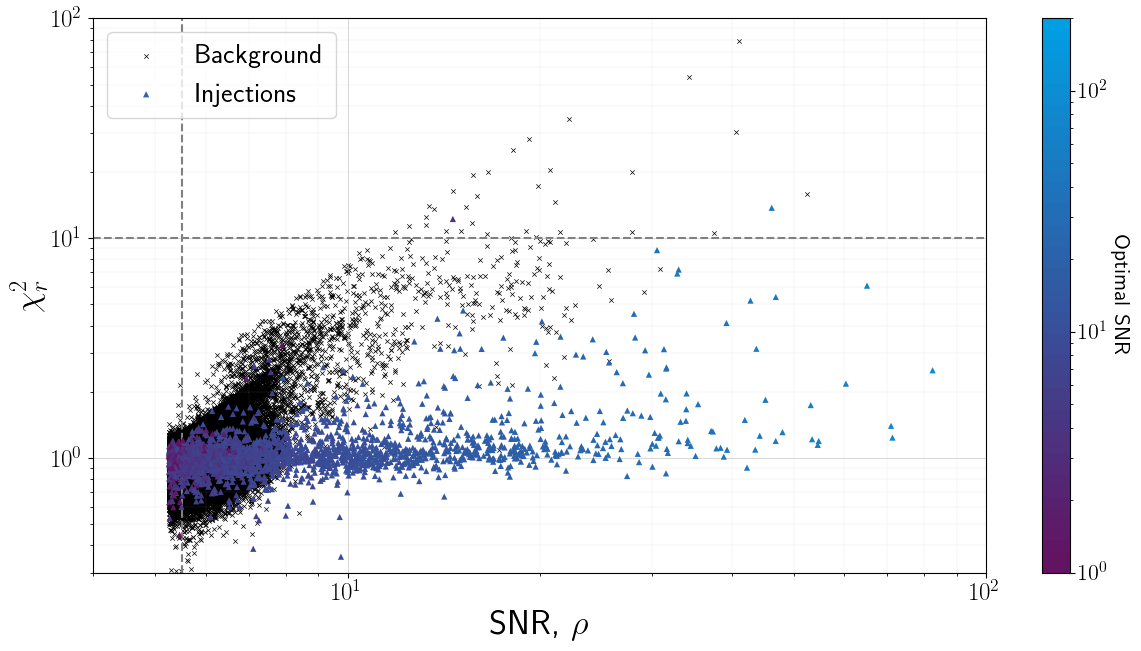}
\end{center}
\caption{\label{fig:snrchi}
Plot of \ac{SNR} and $\chi^2_r$ for background triggers and triggers \removed{from}\inserted{associated with} recovered injections for an analysis of seven days during O3b.
Each point of the scatter plot represents a single-detector trigger, either as found in a coincident injection (coloured triangles), or in the exclusive background of the coincident search (black crosses).
The colour of the injection triggers is based on the optimal \ac{SNR}; the expected \ac{SNR} of the injection if it were to be recovered by an exact match to the waveform it was injected with.
For this figure, the optimal \ac{SNR} can be considered as a measure of the loudness of the injection.
The dashed lines indicate the cuts on single-detector triggers used before ranking statistic computation, where triggers with SNR $\rho<5.5$ or with $\chi^2_r > 10$ are removed from consideration as single-detector triggers.
\inserted{The two particularly high $\chi^2_r$ injections, with $\chi^2_r$ above ten, are signals injected within seconds of rapid bursts of loud glitches.}
}
\end{figure}

\inserted{The \ac{SNR}, $\rho$, of the trigger is generally reduced according to certain properties to produce the \emph{reweighted \ac{SNR}}, $\hat\rho$ \citep{Babak:2012zx}.
One such property is the}\removed{The first cut is on the} reduced $\chi^2$ discriminator of \citep{Allen:2004gu}, $\chi_r^2$, which checks that the \ac{SNR} is accumulated in a way which is consistent with the frequency evolution of the template.
\inserted{Instead of only re-weighting the \ac{SNR}, we decide to impose a stricter criterion, which is a hard cut on the value of $\chi_r^2$.}
By insisting that $\chi^2_r < 10$ we remove the loudest and most obvious glitches.
We see in Figure~\ref{fig:snrchi} that background triggers can have much higher $\chi^2_r$ values than those of (injected) signals, and that nearly all injected signals in this analysis were found using triggers with $\chi^2_r < 10$.

\begin{figure}
\begin{center}
\includegraphics[width=\textwidth]{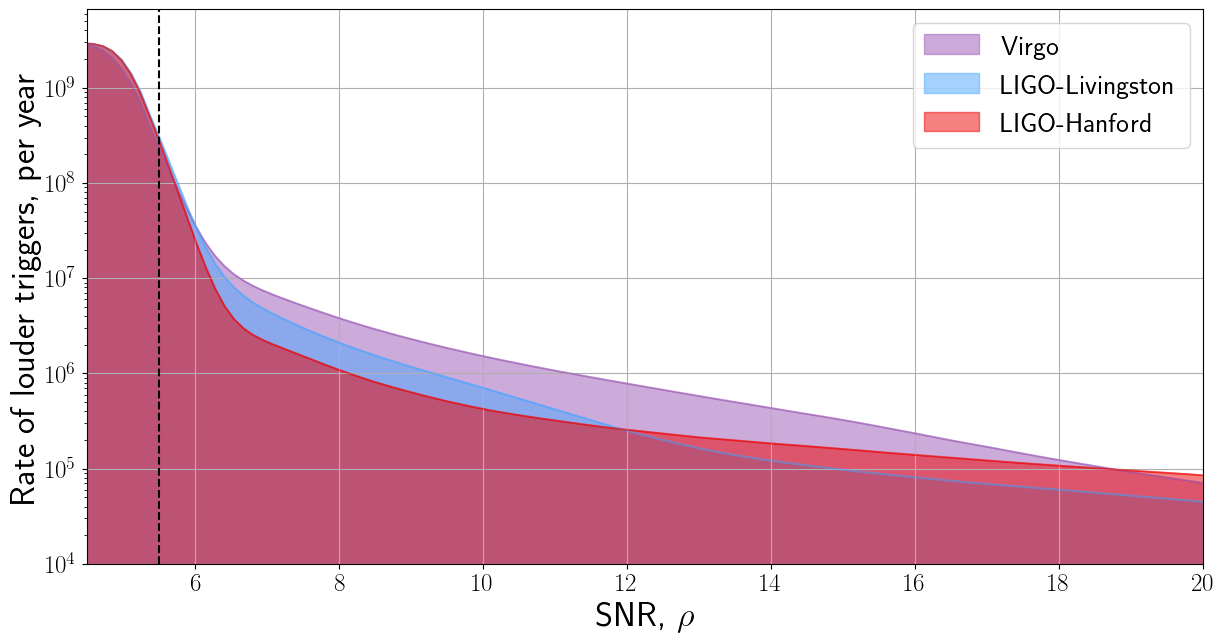}
\end{center}
\caption{\label{fig:snrhist}
Count of triggers above given values of \ac{SNR} for an analysis of seven days during O3b.
	\removed{Triggers are only plotted which meet the $\chi^2_r < 10$ criterion.}
We see that the \ac{SNR} requirement, $\rho > 5.5$, shown by the dashed line, removes most of the quietest triggers, but retains many triggers for use in understanding the background distribution.
}
\end{figure}

The next cut is on the \ac{SNR}, $\rho$, of the signal, which in this work we enforce must be above $5.5$.
Figure~\ref{fig:snrhist} shows the distribution of trigger \ac{SNR} in each detector.
We see that the chosen \ac{SNR} criterion of $\rho>5.5$ is low enough to be well within the noise distribution, but high enough to still remove the majority of triggers.
\removed{The value of the \ac{SNR} cut also needs to be high enough so that the distribution above the cut is not affected by clustering, which causes the reduction in trigger rates below this point.}
\inserted{The triggers are clustered, which means that the trigger with the maximum reweighted \ac{SNR} within a certain window is kept, and so we need to choose a \ac{SNR} cut high enough to not be within the region affected by this clustering.}
Though the triggers just above the \ac{SNR} cut-off are still very unlikely to come from signals, they are still informative for the distribution of the single-detector ranking statistic, and so are useful for later stages of the extrapolation of background.

\subsection{Ranking for false alarm rate calculation}

The ranking statistic we use to compare events is based on the ratio of the expected rate density of signals to an empirically measured noise rate density~\citep{Davies:2020tsx}.
For coincident \removed{triggers}\inserted{events}, this takes the form (Equation 16 of~\citep{Davies:2020tsx});
\begin{equation}
R = - \log A_{N\{d\}} - \sum_d \log r_{di}(\hat{\rho}) + \log p(\vec{\Omega}|S) - \log p(\vec{\Omega}|N) + R_{\sigma,i} .
\label{eq:coincrankingstatistic}
\end{equation}
The constituent parts of the ranking statistic are:
\begin{itemize}
\item $A_{N\{d\}}$ is the allowed ($\left(N_{\textrm{detectors}}-1\right)$-dimensional) time window for coincidences in each detector combination.
\item $r_{di}(\hat{\rho})$ is the \removed{expected}\inserted{measured} rate density of triggers in template $i$ and detector $d$ at reweighted \ac{SNR} $\hat{\rho}$~\citep{Nitz:2017svb}.
\inserted{The measured rate density is that of all triggers, rather than of noise triggers only, and so could include a slight bias from the inclusion of signal triggers, this bias will be naturally very small as the vast majority of triggers will come from noise.
To mitigate this bias, we remove triggers within a 0.1s window of the highest $\hat\rho$ triggers before calculation of the trigger rate density.}
\item $\p(\vec{\Omega}|S)$ is the probability of a signal having the extrinsic parameters, $\vec\Omega$, of the \removed{trigger}\inserted{event} (time difference, phase difference, amplitude ratio) given by prior histograms obtained through a Monte Carlo calculation.
\item $\p(\vec{\Omega}|N)$ is the same probability given a noise distribution, which is assumed to be constant.
\item $R_{\sigma,i}$ is (the log of) network sensitive volume for a given template and triggered detector network, which is proportional to the expected rate of signals, normalised compared to a reference network.
{$R_{\sigma,i} \equiv 3 \left( \log \sigma_{{\rm min}, i} - \log \overline{\sigma}_{HL, i} \right)$} \removed{where $\sigma_{{\rm min}, i}$ is the minimum sensitivity to template $i$ over detectors in the triggered network, and $\overline{\sigma}_{HL, i}$ is the reference sensitivity, for which we use the median network sensitivity to that template in the \ac{HL} detector network.}
\inserted{where $\sigma_i$ is the expected \ac{SNR} of a signal with the same parameters as template $i$ directly overhead at a distance of 1\,Mpc~\citep{Brown:2004vh}.
$\sigma_{{\rm min}, i}$ is where we take the minimum over detectors in the triggered network, and $\overline{\sigma}_{HL, i}$ is the reference sensitivity, for which we use the minimum over detectors of the median $\sigma$ for triggers from template $i$ in the \ac{HL} detector network.}
The reference sensitivity is kept the same for all detector combinations in order to ensure that the ranking statistic is comparable over different triggered network configurations.
\end{itemize}

For the single-detector \removed{trigger}\inserted{event} ranking statistic, we remove all terms which come from the coincident nature of the event.
We therefore remove $A_{N\{d\}}$, $\p(\vec{\Omega}|S)$, and $\p(\vec{\Omega}|N)$ and the ranking statistic becomes
\begin{equation}
	R = - \log r_{di}(\hat{\rho}) + R_{\sigma,i}.
\label{eq:snglsrankingstatistic}
\end{equation}

Outside of the differences in sensitive distance, the ranking statistic is therefore entirely dependent on the \removed{calculated expected}\inserted{measured} rate density of \removed{noise }triggers in each template and detector.

The \removed{expected noise}\inserted{measured} trigger rate for each template and detector, $r_{di}(\hat{\rho})$, is modelled as an exponential decay of the reweighted \ac{SNR} $\hat{\rho}$ of triggers \citep{Nitz:2017svb}; 
\begin{equation}
	r_{di}(\hat{\rho}) = N_{di} \alpha_{di} \exp \left[ - \alpha_{di} (\hat{\rho} - \hat{\rho}_\textrm{thresh} ) \right],
\end{equation}
where $N_{di}$ is the number of triggers in the template with $\hat{\rho} > \hat{\rho}_\textrm{thresh}$, $\hat{\rho}_\textrm{thresh}$ is a threshold value of reweighted \ac{SNR}, and $\alpha_{di}$ is the exponential decay constant.

We use a maximum-likelihood procedure to fit the $\alpha_{di}$ and $N_{di}$ parameters, which are then smoothed over nearby templates.
The smoothed $\alpha_{di}$ and $N_{di}$ are then used to \removed{calculate the expected}\inserted{measure the} rate of triggers at that value of $\hat\rho$.

The smoothing is useful for templates where we have very few triggers with $\hat\rho > \hat{\rho}_\textrm{thresh}$, and we gain more triggers to avoid problems with small sample-size statistics.
However we can face occasional problems with this procedure.

The most problematic situation for this procedure would be where a set of detector artefacts trigger many times in a specific template, but not in nearby templates, for example this can be the case for scattered-light artefacts \citep{LIGO:2020zwl,Accadia:2010zzb}.

\begin{figure}
\begin{center}
\hfill
\includegraphics[width=0.49\textwidth]{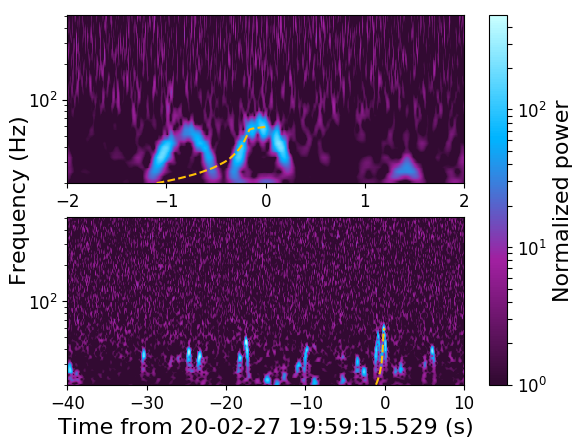}
\hfill
\includegraphics[width=0.49\textwidth]{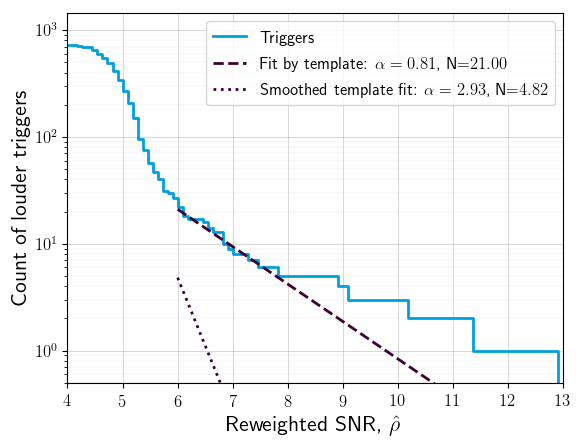}
\hfill
\end{center}
\caption{\label{fig:trig_fits_template}
Example of problematic smoothing of trigger distribution fits (right), and a loud trigger from this template (left).
We see that the triggers in the template are fit with a falling exponential with fit coefficient $\alpha=0.81$, and there are 21 triggers with reweighted \ac{SNR} above threshold.
However once this is smoothed over local templates, the exponential fit steepens to $\alpha=2.93$ and there are 4.82 triggers above threshold per template.
As a result, the \removed{expected }rate of triggers at the \ac{SNR} of the loudest events falls by many orders of magnitude, boosting the ranking of these events.
The example trigger has reweighted \ac{SNR} 22.5, which would have had a\removed{n expected}\inserted{ trigger} count density of around $2.7 \times 10^{-5}$, which changes to a\removed{n expected} count density of $1.4\times 10^{-20}$ with the smoothed fit parameters.
N.B. The example trigger is removed from the triggers used for fitting by a process which removes the loudest triggers in the search.
This example trigger comes from scattered light in the detector.
}
\end{figure}

\begin{figure}
\begin{center}
\hfill
\includegraphics[width=0.49\textwidth]{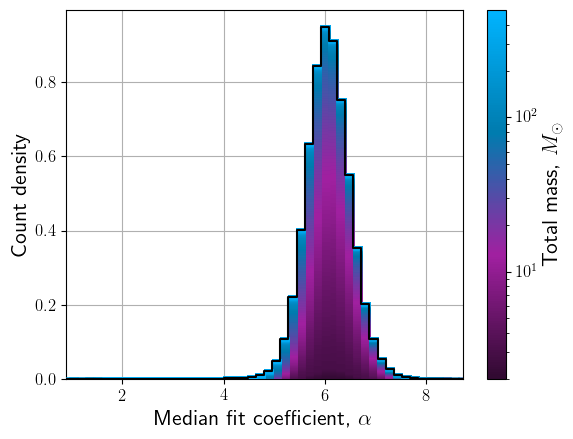}
\hfill
\includegraphics[width=0.49\textwidth]{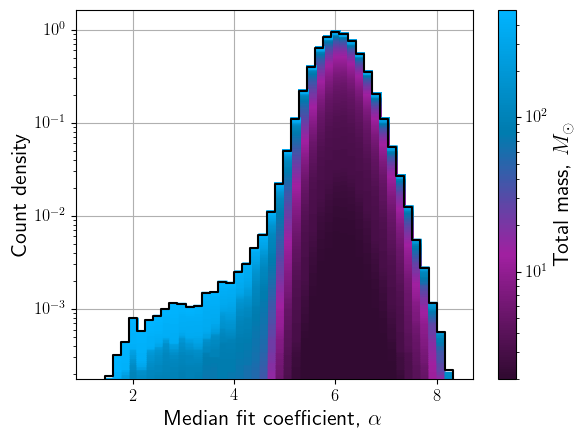}
\hfill
\end{center}
\caption{\label{fig:fit_coeff_med}
\inserted{A histogram of fit coefficients from LIGO-Hanford during the \ac{O3} \PYCBCBROAD analysis, left, and with logarithmic count density plotted, right.
We use the median of the fit coefficient in each template from different analyses over the course of the observing run in order to remove significant outliers and cases where templates had no triggers with reweighted \ac{SNR} above threshold.
We see that the peak of the distribution is around six.
The color of the histogram indicates the total mass of the template which contributed to the bin of the histogram, and we see that many of the higher-mass templates have low fit coefficients; this comes from the relatively short high-mass templates matching well to glitches and therefore having increased trigger rates at higher \ac{SNR}.
The lower-mass templates generally have a well-constrained distribution of fit coefficients around the mode.
}}
\end{figure}

The template fit in Figure~\ref{fig:trig_fits_template} shows an example of this, with a slow exponential decay in this template.
However when it is smoothed by surrounding templates, the expected exponential decay is much steeper, and the calculated rate of \removed{events}\inserted{triggers} at this reweighted \ac{SNR} would be much lower than it should be.
Given Gaussian noise, we would expect the decay in each template to follow an exponential decay distribution as a function of reweighted \ac{SNR} with decay constant of around 6\footnote{As the signal consistency tests we perform are one-sided, and only penalise rather than boost the \ac{SNR} in reweighted \ac{SNR} calculations, the exponential decay constant ends up being just below 6.}\inserted{, as seen in Figure~\ref{fig:fit_coeff_med}}.

In order to remove the effect of over-smoothing where it results in overly optimistic rate estimates, we make our final cut on triggers in templates with an exponential decay constant $\alpha$ below 2.5 before the smoothing is applied.
This has the effect of removing a relatively small fraction of the bank in most cases \inserted{, as seen in Figure~\ref{fig:fit_coeff_med} and Table~\ref{tab:n_removed_alphacut}.}
Table~\ref{tab:n_removed_alphacut} shows the number of removed templates for each search\inserted{, with summary values over different chunks of analysis.
The observing runs are split into shorter sections, which we call chunks, for analysis during an offline search, and these chunks are analysed independently of one another to allow for changes in detector sensitivity and data quality throughout the observing run.}

This cut means that the example trigger shown in Figure~\ref{fig:trig_fits_template} from a template with $\alpha=0.81$ would not be used further in the analysis.

\begin{table}
\begin{center}
\begin{tabular}[htbp]{lccc}
\hiderowcolors
& & \PYCBCBROAD & \PYCBCBBH \\
\hline
\rowcolor{lightgray} & Maximum & 
	\FITOVERREMOVEDTEMPLATESBROADHMAX (\FITOVERREMOVEDTEMPLATESBROADHMAXPERCENT) &
	\FITOVERREMOVEDTEMPLATESBBHHMAX (\FITOVERREMOVEDTEMPLATESBBHHMAXPERCENT) \\
\rowcolor{lightgray}\multirow{-2}{*}{LIGO-Hanford} & Median & \FITOVERREMOVEDTEMPLATESBROADHMEDIAN (\FITOVERREMOVEDTEMPLATESBROADHMEDIANPERCENT) &
	\FITOVERREMOVEDTEMPLATESBBHHMEDIAN (\FITOVERREMOVEDTEMPLATESBBHHMEDIANPERCENT) \\
\rowcolor{white} & Maximum &
	\FITOVERREMOVEDTEMPLATESBROADLMAX (\FITOVERREMOVEDTEMPLATESBROADLMAXPERCENT) &
	\FITOVERREMOVEDTEMPLATESBBHLMAX (\FITOVERREMOVEDTEMPLATESBBHLMAXPERCENT) \\
\rowcolor{white}\multirow{-2}{*}{LIGO-Livingston} & Median & \FITOVERREMOVEDTEMPLATESBROADLMEDIAN (\FITOVERREMOVEDTEMPLATESBROADLMEDIANPERCENT) &
        \FITOVERREMOVEDTEMPLATESBBHLMEDIAN (\FITOVERREMOVEDTEMPLATESBBHLMEDIANPERCENT) \\
\rowcolor{lightgray} & Maximum &
	\FITOVERREMOVEDTEMPLATESBROADVMAX (\FITOVERREMOVEDTEMPLATESBROADVMAXPERCENT)  &
	\FITOVERREMOVEDTEMPLATESBBHVMAX (\FITOVERREMOVEDTEMPLATESBBHVMAXPERCENT) \\
\rowcolor{lightgray}\multirow{-2}{*}{Virgo} & Median & \FITOVERREMOVEDTEMPLATESBROADVMEDIAN (\FITOVERREMOVEDTEMPLATESBROADVMEDIANPERCENT) 
	&
        \FITOVERREMOVEDTEMPLATESBBHVMEDIAN (\FITOVERREMOVEDTEMPLATESBBHVMEDIANPERCENT) \\
\end{tabular}
\end{center}
\caption{\label{tab:n_removed_alphacut}
Number and fraction of templates which had their triggers removed from single-detector event analysis in each detector for the chunks of data analysed by the \PYCBCBROAD and \PYCBCBBH searches.
We present the maximum number and the median, showing worst-case and usual scenarios.
We expect a small number of templates to be removed due to Gaussian noise fluctuations, dependent on the threshold \ac{SNR} value, which we use $\hat{\rho}_\textrm{thresh} = 6$, and proportional to the number of templates and the amount of time analysed.
The \PYCBCBROAD search used \NUMTEMPLATESBROAD templates in one week chunks, and the the \PYCBCBBH search used \NUMTEMPLATESBBH templates in chunks of around a month, explaining the many fewer affected templates. 
}
\end{table}

\subsection{Extrapolation of false alarm rate}
In order to extrapolate the false alarm rate beyond the limit of one event per live time of the search, we fit the number of single-detector events above a particular ranking statistic to a falling exponential.
This is similar to the falling exponential used to estimate the rate of triggers at a particular reweighted \ac{SNR} described above.

\begin{figure}
\begin{center}
\includegraphics[width=0.9\textwidth]{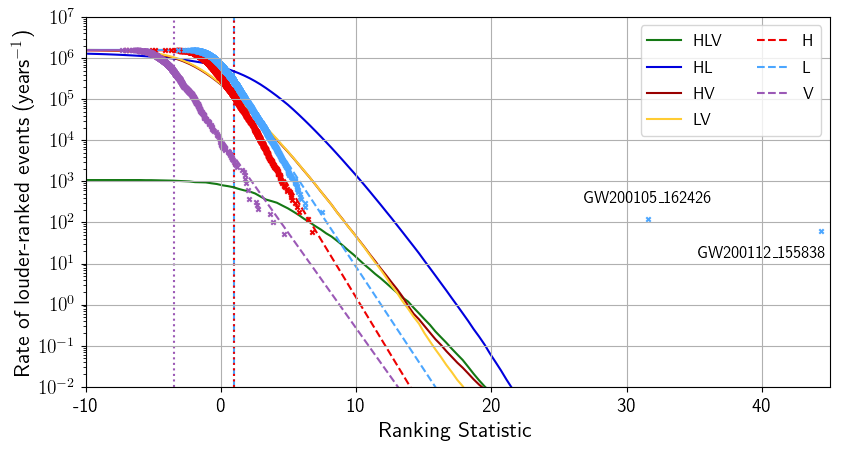}
\end{center}
\caption{\label{fig:chunk31_stat_vs_far}
Rate of louder-ranked events in the exclusive background vs ranking statistic for the \PYCBCBROAD analysis of data between 2020-01-04 17:06:58 and 2020-01-13 10:28:01.
For single-detector events we see the extrapolated background \removed{beyond the limit of the ranked events}\inserted{rate above the threshold ranking statistic}, and \removed{foreground }events with the rate of equal-or-louder ranked events as scatter points.
The lower thresholds for inclusion in the exponential fit are shown as dashed or dotted vertical lines at 1 for LIGO-Hanford and LIGO-Livingston, and at -3.5 for Virgo.
The \ac{FAR} for single-detector events is calculated by finding the extrapolated rate of louder events, the dotted line, at the ranking statistic of the event.
For contrast, we have plotted scatter points which would be the \ac{FAR} when counting equal-or-higher ranked events.
\inserted{It may seem somewhat counter-intuitive for the coincident events to have higher \acp{FAR}, however this is as the distribution of single-detector event ranking statistic peaks around lower values, as the ranking statistic is lower given that we believe they are less likely to be real events.
A single-detector event with a high ranking statistic comparable to those found by coincident searches is therefore extremely rare and has a low \ac{FAR}.}
}
\end{figure}

We fit this exponential for events above a certain ranking statistic threshold.
This threshold must be chosen carefully; we must ensure that enough events are used in the fitting to obtain a confident event rate and exponential decay constant, and balance this consideration against avoiding the effects of clustering the \removed{triggers}\inserted{events} at lower ranking statistic.
For example, in our analysis of \ac{O3} data in Section~\ref{sec:o3_results}, we use fit thresholds of 1 for \ac{H} and \ac{L} events, and of -3.5 for \ac{V} events, as seen in Figure~\ref{fig:chunk31_stat_vs_far}; this different threshold is mainly due to the sensitivity term $R_{\sigma, i}$ being much lower in Virgo than in the LIGO detectors, down-ranking the candidate events.
In order to obtain accurate \ac{FAR} estimates for lower-significance candidate events, we revert to counting the number of louder-ranked events below this threshold.
\inserted{We do not use an explicit upper threshold on the ranking statistic for inclusion in the fit, but any triggers within 0.1s of a (coincident or single-detector) event with \ac{FAR} below 33.3 per year are removed from the fits.}

By using the extrapolation of the ranking statistic--false alarm rate relationship, we are now able to estimate significance beyond the limit of one per live time.
The extrapolation method we have detailed here is for single-detector events only, and so we now consider it as part of a wider search including coincident events.

\section{Using the estimated significance in a wider search context}
\label{sec:combining-far}

Once we have calculated the \acp{FAR} for the events in their detector combinations, these are combined with results from all possible combinations.
The best-ranked event from any detector combination at a given time is used, and so through the `look elsewhere' effect, we will find more false alarms than we would looking in a single detector combination.

Often this effect would be included through multiplication by a trials factor, but that would not be appropriate here given the vastly different rate of false alarms in the different possible combinations of detectors.
We follow the same method as~\citep{Davies:2020tsx}, and obtain an overall \ac{FAR} by summing \acp{FAR} at the ranking statistic of the candidate event in all available combinations at that time.
This means that a candidate event in \ac{HLV} time would have the \ac{FAR} at its ranking statistic combined from the \ac{H}, \ac{L}, \ac{V}, \ac{HL}, \ac{HV}, \ac{LV} and \ac{HLV} \inserted{\acp{FAR}}\removed{backgrounds}, but a candidate event in \ac{LV} time would only have the \ac{LV}, \ac{L} and \ac{V} \acp{FAR} added together.

We consider what happens in this procedure in the context of single-detector \removed{triggers}\inserted{events}.

For a \ac{FAR} \removed{is }based on an exponential fit with no limits at high statistic, highly-ranked events will get negligible \ac{FAR}.
At high statistic, the coincident backgrounds are much higher than the \inserted{estimated} backgrounds for single-detector events, as seen in Figure~\ref{fig:chunk31_stat_vs_far}.
At very high statistic, if a coincident event is ranked louder than all background events, its \ac{FAR} is set to one per background time; the amount of time which the time-shifted background explores.
The background time depends on the analysis time and the number of time shifts, and is of the order of tens of millenia for both \PYCBCBROAD and \PYCBCBBH.

An event with very low \ac{FAR} is therefore naturally mitigated by the combination with the coincident backgrounds when other detector combinations are available.
Additionally, the single-detector backgrounds do not have a large effect on the significance estimates of confidently-detected coincident events.

For single-detector events in single-detector time, we set a limit on \ac{FAR} corresponding to the largest available coincident background time, to prevent unbelievably small \acp{FAR}.
This cut-off only affects the very loudest events, and once we get to low values of \ac{FAR}, we do not require more confidence, as we are already almost certain that the event is real.
In catalogue papers, \acp{FAR} are often quoted as being below a particular cutoff; in 3-OGC and 4-OGC, this was a \ac{FAR} of one per hundred years, and in GWTC-2.1 and GWTC-3, this was one per $10^{5}$ years.

By combining the extrapolated \ac{FAR} of single-detector events with the \ac{FAR} from coincident events at the same ranking statistic, we have included the single-detector events in the \PYCBC offline search, ensuring that the results remain sensible for all events.

\section{Results from the third LIGO observing run}
\label{sec:o3_results}
In order to test our method, we compare to \removed{the}\inserted{a} coincident-only \PYCBC search, similar to the analyses used in GWTC-2.1 from \ac{O3a} \citep{LIGOScientific:2021usb} and GWTC-3 from \ac{O3b}~\citep{LIGOScientific:2021djp}.

In each of these papers, there were two analyses, \PYCBCBROAD and \PYCBCBBH, which differ slightly from one another.
The \PYCBCBBH analysis uses a much smaller template bank focussed on the stellar-mass black hole region in which we have found many events already, whereas \PYCBCBROAD is a much wider analysis, including \ac{BNS} and \ac{NSBH} events, and is able to recover events which may be outside of the parameter space in which we have already seen events.

In addition, the \PYCBCBBH analysis uses an explicit model of the black hole mass population, and so the ranking statistic has an additional term of $-\frac{11}{3}\log(\mathcal{M}_i / \mathcal{M}_\textrm{ref})$, where $\mathcal{M}_i$ is the chirp mass~\cite{Peters:1964zz,Blanchet:1995ez} of the triggered template and $\mathcal{M}_\textrm{ref}=20\Msun$ is a reference chirp mass~\citep{Dent:2013cva,Nitz:2019hdf}.
This term should be truncated above $\mathcal{M} = 40\Msun$, however due to an error in implementation this was not applied in the GWTC-2.1 and GWTC-3 results~\citep{LIGOScientific:2021djp}.
This bug up-ranked many triggers, and as single-detector events are particularly sensitive to issues in the ranking statistic, the bug was corrected for this work in all our \PYCBCBBH{} analyses.

In the \ac{GWTC} papers, in order to prevent background contamination by single-detector events, \removed{triggers were removed from the background if they were}\inserted{background events were removed if they contained triggers} within $\pm0.1$\,s of events found by other searches with \ac{FAR} below one per hundred years, provided that they did not form coincidences in the \PYCBC searches at any significance.
This one-per-hundred-years limit was designed to match the hierarchical removal stage of the \PYCBC analysis.

In GWTC-2.1, the \removed{removed }events\inserted{ used for trigger removal} came from the GWTC-2 catalog \citep{LIGOScientific:2020ibl}.
For GWTC-3, the \removed{removed }events\inserted{ used for trigger removal} came from events found by the \GSTLAL search in that paper.

In 4-OGC~\citep{Nitz:2021zwj}, the \pastro{} calculation for candidate\inserted{ event}s in a single detector utilised only the background when both LIGO detectors are observing in order to minimise possible signal contamination.
Coincident events do not have the single-detector signals removed from the background, as the primary figure of merit for that work is \pastro{}, which remains high for \ac{BBH} events at the point where background contamination becomes an issue.
The \ac{FAR} is a more suitable figure of merit for events where the signal distribution is less understood, and we see in 4-OGC that the single-detector \ac{BNS} and \ac{NSBH} events \FULLNAME{GW190425} and \FULLNAME{GW200105} are recovered with $\pastro{} \sim 0.5$, limited by the signal rates in that region.

To compare coincident-only analyses with the singles-included search, we re-analysed the data using the same method as in GWTC-2.1 and GWTC-3, but did so without removing the single-detector event triggers listed above from the background, this is denoted as the \emph{coincident-only} search.
This means that the results we compare to are not as accurate or optimistic as the results from those analyses, however they are more representative of independent offline \PYCBC analyses than as presented in the GWTC catalogs.

In addition to the non-removal of single detector events from the background, the results do not exactly match the results in the GWTC papers for two reasons; the GWTC papers use probability of astrophysical origin \pastro{} as the threshold for inclusion, and the \PYCBC results in those papers state the inclusive \ac{FAR}, but the \ac{FAR} used in this paper is exclusive.

The inclusive false alarm rate is defined under the assumption that the event of interest, and any\removed{thing}\inserted{ event} ranked lower is noise, and so all triggers at the time of the event are included in the background.
The \ac{FAR} given by an inclusive background is estimated by successively removing the triggers within a 0.1\,s window of the \inserted{foreground }events in descending order of \ac{FAR}, calculating \ac{FAR} for each event before removing \removed{its}\inserted{nearby} triggers from the background.
We do this for all events with \ac{FAR} below one per hundred years.
The exclusive false alarm rate is built under the assumption that all the events we find are signals, and so all triggers at the time of events are excluded from the background.
We therefore remove all triggers within a 0.1\,s window of any event with \ac{FAR} less than one per 0.03\,years for this calculation.

The inclusive \removed{background }\ac{FAR} is not useful for single-detector signals, as it will always be less than the live-time of the search.
One could assume a distribution of signal ranking statistics in order to extrapolate the inclusive \ac{FAR} beyond this limit \citep{Callister:2017urp,Nitz:2020naa}, but in this work, we do not assume any signal distribution properties.
The analyses are split into chunks with live times of around $\sim$ one week for \PYCBCBROAD analysis and $\sim$ one month for the \PYCBCBBH analysis, and so the un-extrapolated inclusive \ac{FAR} would always be found to be insignificant.

Tables~\ref{tab:results_o3a} and~\ref{tab:results_o3b} give the events found in O3a and O3b respectively using this search technique, compared to the results of the coincident-only analyses.

\begin{landscape}
\begingroup
\makeatletter
\begin{table}[]
\scriptsize
\begin{center}
\begin{tabular}{l c cccc ccccc}
\hline
\hiderowcolors
\multicolumn{1}{c}{\multirow{4}{*}{Event}} & \multirow{2}{*}{Instruments} & \multicolumn{4}{c|}{Search including single-detector events} & \multicolumn{4}{c}{Coincident-only search} \\
 & & \multicolumn{2}{c|}{\PYCBCBROAD} & \multicolumn{2}{c|}{\PYCBCBBH} & \multicolumn{2}{c|}{\PYCBCBROAD} & \multicolumn{2}{c}{\PYCBCBBH} \\
 & \multirow{2}{*}{Active, Triggered} &  FAR   &  Network  &  FAR   & Network  &  FAR   & Network  &  FAR &  Network  \\
 & & (years$^{-1}$) & SNR & (years$^{-1}$) & SNR & (years$^{-1}$) & SNR & (years$^{-1}$) & SNR \\
\hline
\showrowcolors
{\EVENTNAMEFONT{GW190814}\FULLNAME{GW190814}} &
                \ACTIVEINSTRUMENTS{GW190814}, \ALLINSTRUMENTS{GW190814} &
                $ \SNGLBROADASSOC{GW190814}  \SNGLBROADFAR{GW190814} $ &
                $ \SNGLBROADASSOC{GW190814} \SUBSETINSTRUMENTSSNGLBROAD{GW190814} \SNGLBROADNETWORKSNR{GW190814} $ &
                $ \SNGLBBHASSOC{GW190814}  \SNGLBBHFAR{GW190814} $ &
                $ \SNGLBBHASSOC{GW190814} \SUBSETINSTRUMENTSSNGLBBH{GW190814} \SNGLBBHNETWORKSNR{GW190814} $ &
                $ \COINCBROADASSOC{GW190814} \COINCBROADFAR{GW190814} $ &
                $ \COINCBROADASSOC{GW190814} \SUBSETINSTRUMENTSCOINCBROAD{GW190814} \COINCBROADNETWORKSNR{GW190814} $ &
                $ \COINCBBHASSOC{GW190814} \COINCBBHFAR{GW190814} $ &
                $ \COINCBBHASSOC{GW190814} \SUBSETINSTRUMENTSCOINCBBH{GW190814} \COINCBBHNETWORKSNR{GW190814} $ \\
\end{tabular}
\end{center}
\caption{
\label{tab:results_o3a}
Results from the \PYCBCBROAD and \PYCBCBBH analyses with singles-included and coincident-only searches in O3a.
We list the \acp{FAR} and network \acp{SNR} from each event for all analyses.
Included are events with \ac{FAR} less than two per year in any analysis, except those for which the \acp{FAR} do not differ between the searches; these events are included in Table~\ref{tab:other_results} in~\ref{app:further_results}.
Events with names in bold were found according to the \ac{FAR} criterion by this work, but did not reach the same criterion in the coincident analyses by the same search.
Events with network \ac{SNR} and \ac{FAR} given in italics are included as they are at the same time as one which meets the criterion for inclusion.
Instruments are given according to the initials of the detectors involved, \ac{H}, \ac{L} or \ac{V}, e.g. a \ac{HL} event comes from LIGO-Hanford and LIGO-Livingston, but not Virgo.
In some cases, the set of instruments which contributed to the events differs between the analyses; for these events, we have listed the largest group of instruments which triggered in the Instruments column.
Where a subset of the instruments listed in the instruments column triggered to form the event, this is indicated with a $^\dagger$.
}

\end{table}
\makeatother
\endgroup

\end{landscape}

\begin{landscape}
\begingroup
\makeatletter
\begin{table}[]
\scriptsize
\begin{center}
\begin{tabular}{l c cccc ccccc}
\hline
\hiderowcolors
\multicolumn{1}{c}{\multirow{4}{*}{Event}} & \multirow{2}{*}{Instruments} & \multicolumn{4}{c|}{Search including single-detector events} & \multicolumn{4}{c}{Coincident-only search} \\
 & & \multicolumn{2}{c|}{\PYCBCBROAD} & \multicolumn{2}{c|}{\PYCBCBBH} & \multicolumn{2}{c|}{\PYCBCBROAD} & \multicolumn{2}{c}{\PYCBCBBH} \\
 & \multirow{2}{*}{Active, Triggered} &  FAR   &  Network  &  FAR   & Network  &  FAR   & Network  &  FAR &  Network  \\
 & & (years$^{-1}$) & SNR & (years$^{-1}$) & SNR & (years$^{-1}$) & SNR & (years$^{-1}$) & SNR \\
\hline
\showrowcolors
{\EVENTNAMEFONT{GW200112}\FULLNAME{GW200112}} &
                \ACTIVEINSTRUMENTS{GW200112}, \ALLINSTRUMENTS{GW200112} &
                $ \SNGLBROADASSOC{GW200112}  \SNGLBROADFAR{GW200112} $ &
                $ \SNGLBROADASSOC{GW200112} \SUBSETINSTRUMENTSSNGLBROAD{GW200112} \SNGLBROADNETWORKSNR{GW200112} $ &
                $ \SNGLBBHASSOC{GW200112}  \SNGLBBHFAR{GW200112} $ &
                $ \SNGLBBHASSOC{GW200112} \SUBSETINSTRUMENTSSNGLBBH{GW200112} \SNGLBBHNETWORKSNR{GW200112} $ &
                $ \COINCBROADASSOC{GW200112} \COINCBROADFAR{GW200112} $ &
                $ \COINCBROADASSOC{GW200112} \SUBSETINSTRUMENTSCOINCBROAD{GW200112} \COINCBROADNETWORKSNR{GW200112} $ &
                $ \COINCBBHASSOC{GW200112} \COINCBBHFAR{GW200112} $ &
                $ \COINCBBHASSOC{GW200112} \SUBSETINSTRUMENTSCOINCBBH{GW200112} \COINCBBHNETWORKSNR{GW200112} $ \\
\end{tabular}
\end{center}
\caption{
\label{tab:results_o3b}
Results from the \PYCBCBROAD and \PYCBCBBH analyses with singles-included and coincident-only searches in O3b.
We list the \acp{FAR} and network \acp{SNR} from each event for all analyses.
The event inclusion criteria and table format are the same as in Table~\ref{tab:results_o3a}.
}
\end{table}
\makeatother
\endgroup

\end{landscape}

In Tables~\ref{tab:results_o3a} and~\ref{tab:results_o3b}, we see that we manage to recover all of the single-detector signals from \acp{GWTC} 2.1 and 3, and all but \FULLNAME{GW190424} from 3-OGC and 4-OGC.
\FULLNAME{GW190424} is a LIGO-Livingston-only event recovered with \pastro{} 0.81 in 3-OGC but is not found with any significance by this anaysis.

Event names encode the event time according to the convention GWYYMMDD\_hhmmss, for example the event \FULLNAME{GW200112} was found at time \STRTIME{GW200112}UTC.
The event \FULLNAME{200218} does not have the `GW' prefix, as it comes from a glitch in the detector, as discussed later.
The network \ac{SNR} is calculated from the sum of squares of \acp{SNR} from the triggers which form the event, and does not account for detectors that did not produce triggers contributing to the event.

The events listed in bold come from a few broad categories of events.
Firstly we see \NUMSNGLEVENTS single-detector events; \NEWSNGLEVENTS, which are assigned significant \ac{FAR} for the first time by \PYCBC searches.

Secondly, there are events which were found by either the \PYCBCBROAD or \PYCBCBBH search in the coincident-only search, but which are found in the other search as a single-detector event; this was the case for \FULLNAME{GW190630}, which is newly found in the \PYCBCBROAD search.
The \ac{FAR} of \FULLNAME{GW190630} is also significantly improved in the \PYCBCBBH search for a different reason, as the \inserted{triggers around the }event \FULLNAME{GW190708} \removed{is}\inserted{are} removed from the background.

The event \FULLNAME{GW190814} is newly found in the \PYCBCBBH search as a single event.
Though the \ac{FAR} of \FULLNAME{GW190814} does not change in the \PYCBCBROAD search, as it is already ranked higher than the loudest background event, the event changes from being an \COINCBROADINSTRUMENTS{GW190814} coincident event with ranking statistic \COINCBROADSTAT{GW190814} to an \SNGLBROADINSTRUMENTS{GW190814} single-detector event with ranking statistic \SNGLBROADSTAT{GW190814}.

\begin{figure*}
	\begin{center}
	\subfloat[\SNGLBROADINSTRUMENTS{GW190425}: \FULLNAME{GW190425}]{\includegraphics[width=0.32\textwidth]{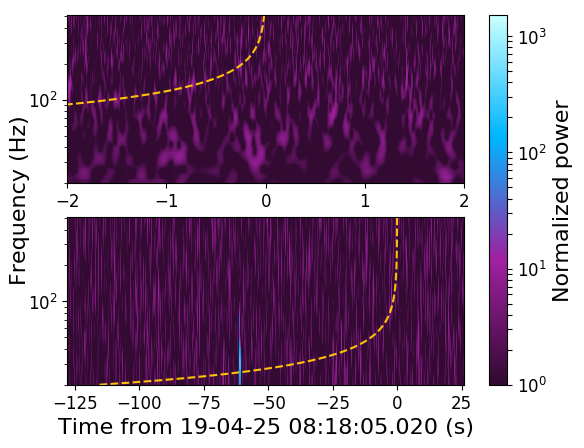}}
	\hfill
	\subfloat[\SNGLBBHINSTRUMENTS{GW190620}: \FULLNAME{GW190620}]{\includegraphics[width=0.32\textwidth]{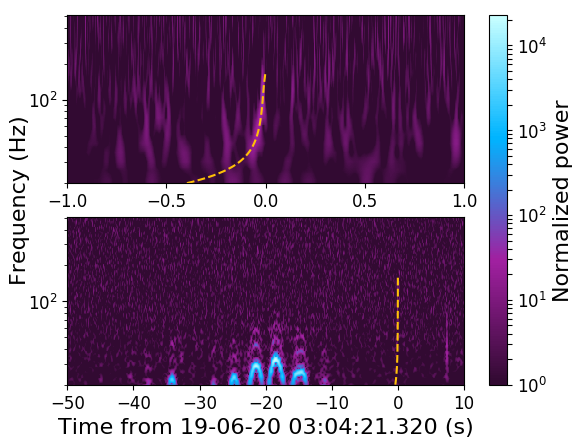}}
	\hfill
	\subfloat[\SNGLBROADINSTRUMENTS{GW190630}: \FULLNAME{GW190630}]{\includegraphics[width=0.32\textwidth]{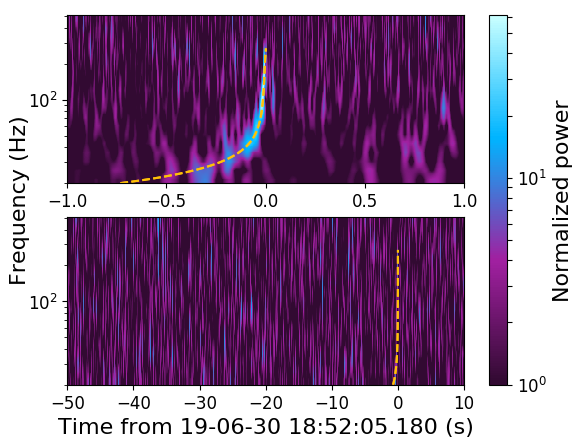}}
	\hfill \\
	\subfloat[\SNGLBROADINSTRUMENTS{GW190708}: \FULLNAME{GW190708}]{\includegraphics[width=0.32\textwidth]{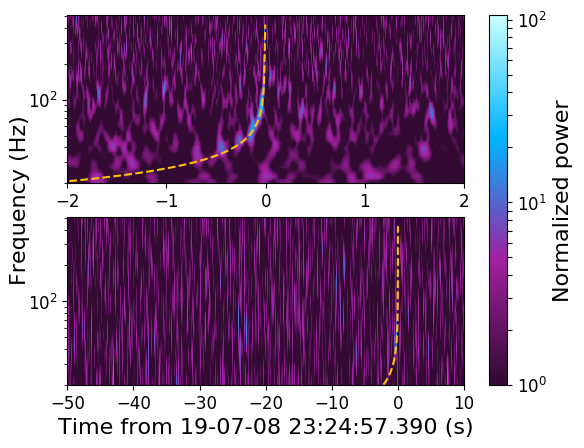}}
	\hfill
	\subfloat[\SNGLBROADINSTRUMENTS{GW190814}: \FULLNAME{GW190814}]{\includegraphics[width=0.32\textwidth]{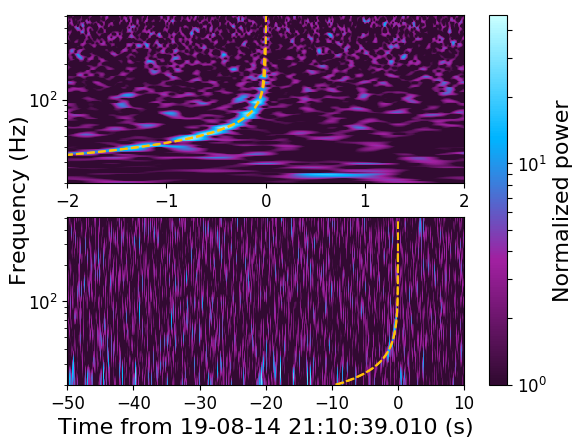}}
	\hfill
	\subfloat[\SNGLBROADINSTRUMENTS{GW190910}: \FULLNAME{GW190910}]{\includegraphics[width=0.32\textwidth]{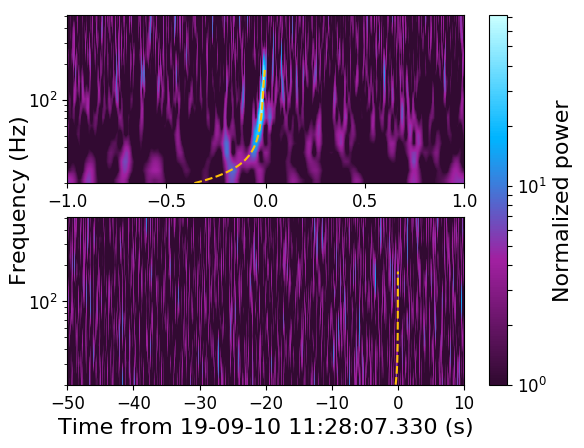}}
	\hfill \\
	\subfloat[\SNGLBROADINSTRUMENTS{GW200105}: \FULLNAME{GW200105}]{\includegraphics[width=0.32\textwidth]{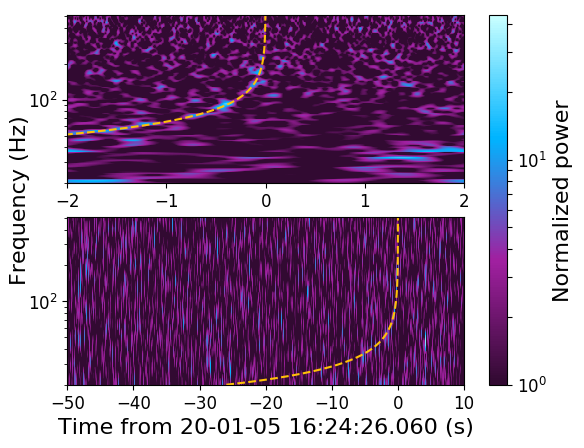}}
	\hfill
	\subfloat[\SNGLBROADINSTRUMENTS{GW200112}: \FULLNAME{GW200112}]{\includegraphics[width=0.32\textwidth]{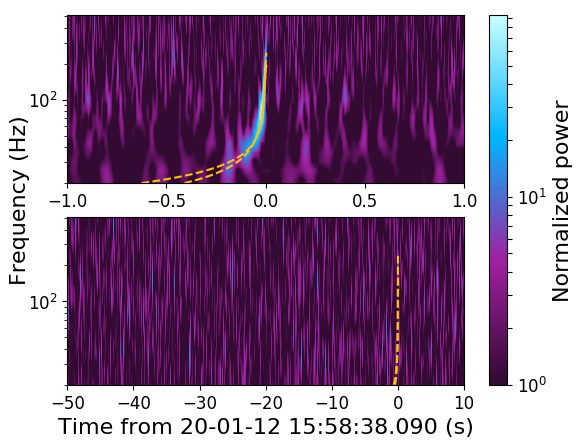}}
	\hfill
		\subfloat[\SNGLBBHINSTRUMENTS{200218}: \FULLNAME{200218}]{\label{fig:200218_qscan}\includegraphics[width=0.32\textwidth]{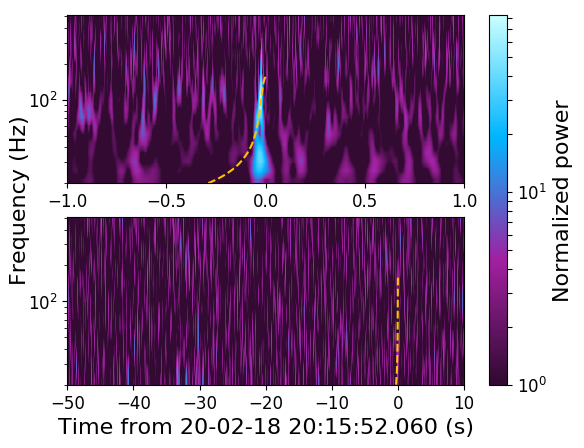}}
	\hfill \\
	\subfloat[\SNGLBBHINSTRUMENTS{GW200302}: \FULLNAME{GW200302}]{\includegraphics[width=0.32\textwidth]{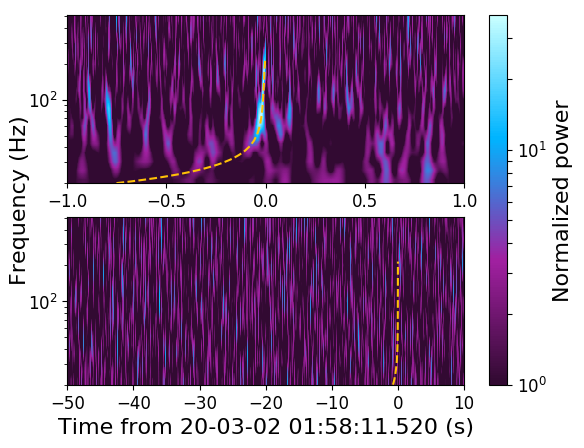}}
	\end{center}
\caption{\label{fig:qscan_plots}
	Time-frequency plots for the single-detector events found by the singles-included analyses.
	The subcaption indicates the detector and which event is plotted.
	The time--frequency track of the template(s) which triggered from the single-detector searches is(are) overlaid.
}
\end{figure*}

Finally we see \FULLNAME{200218} which, as can be seen in Figure~\ref{fig:200218_qscan}, is caused by a glitch in the Virgo detector.
This event has ranking statistic \SNGLBBHSTAT{200218}, which would have \ac{FAR} of over ten per year for LIGO-Hanford and LIGO-Livingston singles, and hundreds per year for any coincident events.
As a result, \FULLNAME{200218}'s significance is \removed{helped}\inserted{boosted} by the fact that it is a Virgo single-detector event in Virgo-only time and does not have backgrounds from other event types suppressing it.

No events lost significance in order to move above the two-per-year threshold for inclusion in the results.

Figure~\ref{fig:qscan_plots} shows time--frequency plots of the events newly-found by this work.
We see that a few of the events are strong enough to be seen by eye in the plots, these are the strong signals we would be concerned about missing in a coincident-only search.

We have seen that using the method described here, we are able to recover signals seen in a single detector, as well as improving our estimates of the significance of other signals by removing the\inserted{ triggers close to} single-detector events from the background estimates\inserted{ of coincident events}.

\section{Search sensitivity}
\label{sec:sensitivity}

We have seen the results on O3 data in Section~\ref{sec:o3_results}, and here we consider injected signals in the data.
By injecting signals from various parts of the \ac{CBC} parameter space, we can assess the change in sensitivity to different signals. 
By comparing which injections were recovered in each search, we can estimate the change in the number of events we would expect to find.
The injections we use are the same as those used in the GWTC-2.1 and GWTC-3 catalogs, and their distributions are described fully in the Appendix of the GWTC-3 paper \citep{LIGOScientific:2021djp}.

First we will discuss the change in the sensitivity of the search, and then discuss the situations which contribute to the change in sensitivity.
To do this, we compare the sensitive volume--time (\VT{}) of each search, which is a measure proportional to the number of signals we would expect to see in a search.

\begin{figure}
\begin{center}
\includegraphics[width=0.49\textwidth]{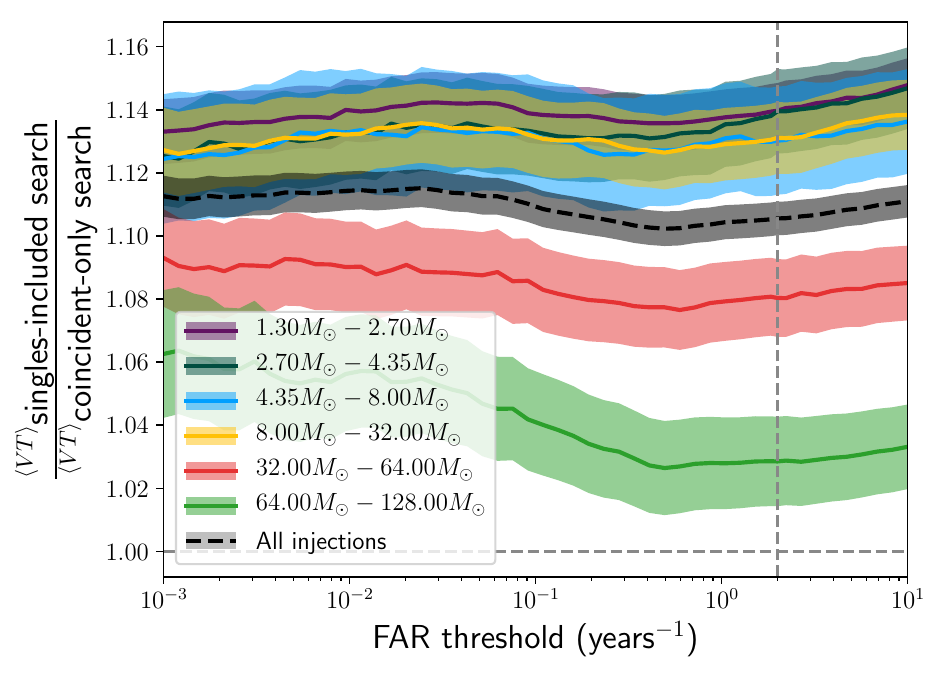}
\hfill
\includegraphics[width=0.49\textwidth]{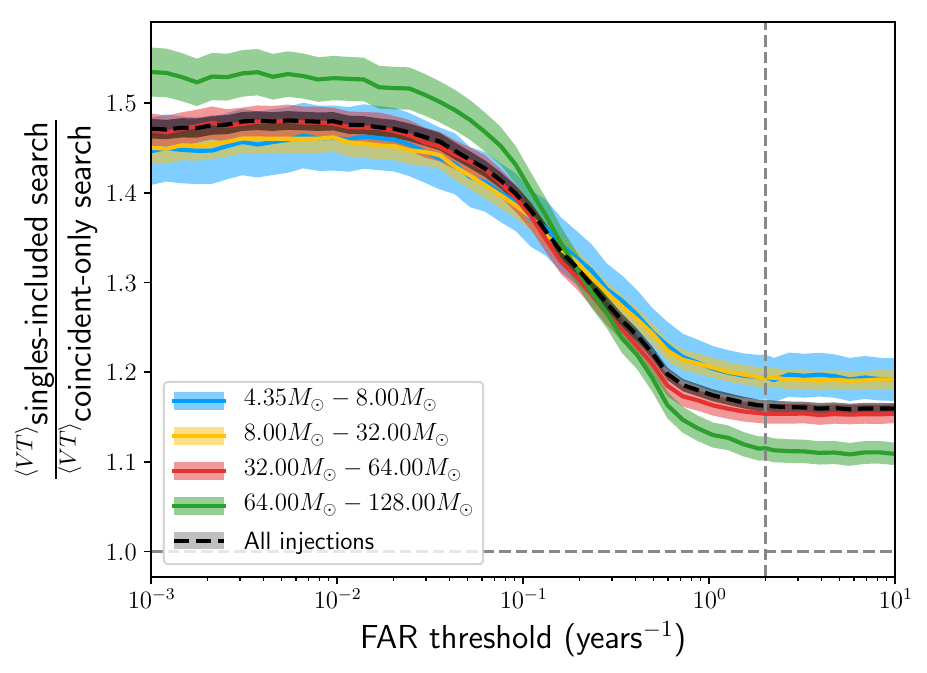}
\end{center}
\caption{\label{fig:vt_ratio}
The ratio of the \VT{} of the singles-included and coincident-only searches at different \ac{FAR} thresholds for the \PYCBCBROAD (left) and \PYCBCBBH (right) analyses.
Injections in different chirp mass bins are denoted by colour, and the \VT{} for the whole population is also included as the black dashed line.
Uncertainty bands are based on Poisson counting uncertainty for the found injections.
}
\end{figure}

Figure~\ref{fig:vt_ratio} shows the ratio of the sensitive volume--time ($R_{\VT}$) of each analysis,where 
\begin{equation}
R_{\VT} = \frac{\VT_\mathrm{singles-included~search}}{\VT_\mathrm{coincident-only~search}},
\label{eqn:vt_ratio}
\end{equation}
which is estimated by counting the number of injected signals and the number of recovered signals as a function of \ac{FAR}.

We use chirp mass ($\mathcal{M}$) bins to show the effect of including single-detector events on the search in different parts of the parameter space.
The bins \VTBINONELIMITS and \VTBINTWOLIMITS match the \ac{BNS} and \ac{NSBH} chirp mass bins used for the \pastro calculations for the \PYCBC searches in \citep{LIGOScientific:2021djp}.

We see that $R_{\VT{}}$ is highest for the lowest-mass bins of injections; this is because the longer-duration waveforms in these bins ensure that the trigger distributions are closer to those from Gaussian noise, as they are not influenced so much by the glitches to which the single-detector search is more susceptible.
At lower \acp{FAR}, we see that the \PYCBCBBH $R_{\VT{}}$ increases significantly, this is due to the background contamination by strong single-detector signals, which given the longer analysis time of the \PYCBCBBH chunks, affect more time (and therefore injections) than the shorter \PYCBCBROAD analyses.

Table~\ref{tab:vt_ratio} shows the \VT{} ratio $R_{\VT}$ for the \PYCBCBROAD and \PYCBCBBH searches at the \ac{FAR} threshold of two per year used in the results of Section~\ref{sec:o3_results}.
We see the results in different mass bins, and averaged for all injections.
For all injections in the \PYCBCBROAD search, we see in increase in sensitive \VT{} by a factor of \VTRATIOBROADALL, and by a factor of \VTRATIOBBHALL in the \PYCBCBBH search.

\begingroup
\makeatletter
\rowcolors{1}{lightgray}{}
\begin{table}[]
\begin{center}
\begin{tabular}{c|cc}
\hline
\hiderowcolors
Chirp Mass bin & \PYCBCBROAD $R_{\VT{}}$ & \PYCBCBBH $R_{\VT{}}$ \\
\hline
\showrowcolors
 \VTBINONELIMITS & \VTRATIOBROADBINONE & \VTRATIOBBHBINONE \\ 
 \VTBINTWOLIMITS & \VTRATIOBROADBINTWO & \VTRATIOBBHBINTWO \\ 
 \VTBINTHREELIMITS & \VTRATIOBROADBINTHREE & \VTRATIOBBHBINTHREE \\ 
 \VTBINFOURLIMITS & \VTRATIOBROADBINFOUR & \VTRATIOBBHBINFOUR \\ 
 \VTBINFIVELIMITS & \VTRATIOBROADBINFIVE & \VTRATIOBBHBINFIVE \\ 
 \VTBINSIXLIMITS & \VTRATIOBROADBINSIX & \VTRATIOBBHBINSIX \\ 

\hline
All injections & \VTRATIOBROADALL & \VTRATIOBBHALL \\
\end{tabular}
\end{center}
\caption{
\label{tab:vt_ratio}
Sensitive \VT{} ratio for the \PYCBCBROAD and \PYCBCBBH searches for signals in various chirp mass bins, and averaged for all injections.
}

\end{table}
\makeatother
\endgroup

\begin{figure}
\begin{center}
	\subfloat[]{\includegraphics[width=0.49\columnwidth]{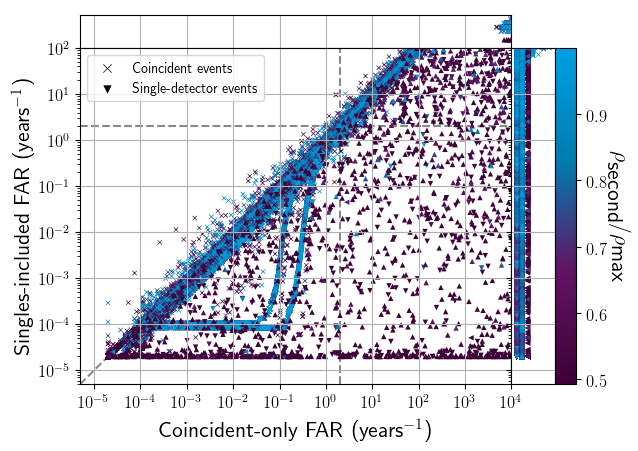}\label{fig:injections_scatter_broad}} \hfill
	\subfloat[]{\includegraphics[width=0.49\columnwidth]{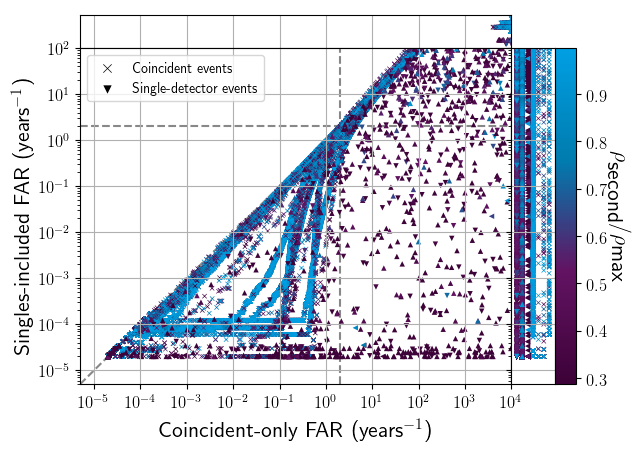}\label{fig:injections_scatter_bbh}}
\end{center}
\caption{\label{fig:injections_scatter}
Recovered \ac{FAR} values for injections from the singles-included analysis compared to the \ac{FAR} of the same injections from the coincident-only analysis for the O3 \PYCBCBROAD search (left) and \PYCBCBBH search (right).
The shape of each scatter point indicates whether the injection was recovered as a coincident or single-detector event in the singles-included analysis.
Each scatter point is coloured according to the ratio of the second largest optimal \ac{SNR} versus the maximum optimal \ac{SNR} over the detectors.
The dashed lines indicate the two-per-year cutoff used to list events in Section~\ref{sec:o3_results}, meaning that events in the lower-right section of the plot would be newly-found by the singles-included search.
We see the effect of removing events from the coincident background, as well as the additional events recovered using the singles-included analysis.
The panels at the side and top of the plot show events which were either completely missed by the search or above the \ac{FAR} limits of the plot.
}
\end{figure}


Figure~\ref{fig:injections_scatter} compares the recovered \ac{FAR} of injections from the coincident-only analysis and singles-included analysis, for the \PYCBCBROAD search (left) and \PYCBCBBH search (right).
We see that though there are a few injections found with better \ac{FAR} in the coincident-only search, most of the injections are found with improved \ac{FAR} by including the single-detector events.
We also see the many newly-found events which were completely missed by the coincident search in the side panel.

A useful metric for assessing whether whether a signal is likely to be found is the optimal \ac{SNR}.
The optimal \ac{SNR} is the \ac{SNR} which would be recovered by an exact-match template in zero noise given the \ac{PSD} at the time of the event.
The detector with the maximum optimal \ac{SNR} would be decisive in a search involving single-detector events, however for coincident-only searches, the decisive optimal \ac{SNR} would be the second largest \ac{SNR} over the set of operating detectors.
We use the ratio of these two optimal \acp{SNR}, $\rho_\textrm{second} / \rho_\textrm{max}$, in Figure~\ref{fig:injections_scatter}, and the biggest improvement is for events with the lowest value of this ratio, with darker scatter points.
Events with a low value of this ratio are loud in one detector, but not in the next-loudest operating detector.

We additionally see arcs of injections with significantly improved recovered \ac{FAR}; these come from the analysis chunks containing loud single-detector events.
For \PYCBCBROAD, we see two arcs, corresponding to the chunk containing \FULLNAME{GW200105} and \FULLNAME{GW200112}, and the one containing \FULLNAME{GW190630}.
The higher \acp{FAR} in these arcs in the coincident-only search come from the presence of loud single-detector events in the background.

\begin{figure}
\begin{center}
\includegraphics[width=0.9\columnwidth]{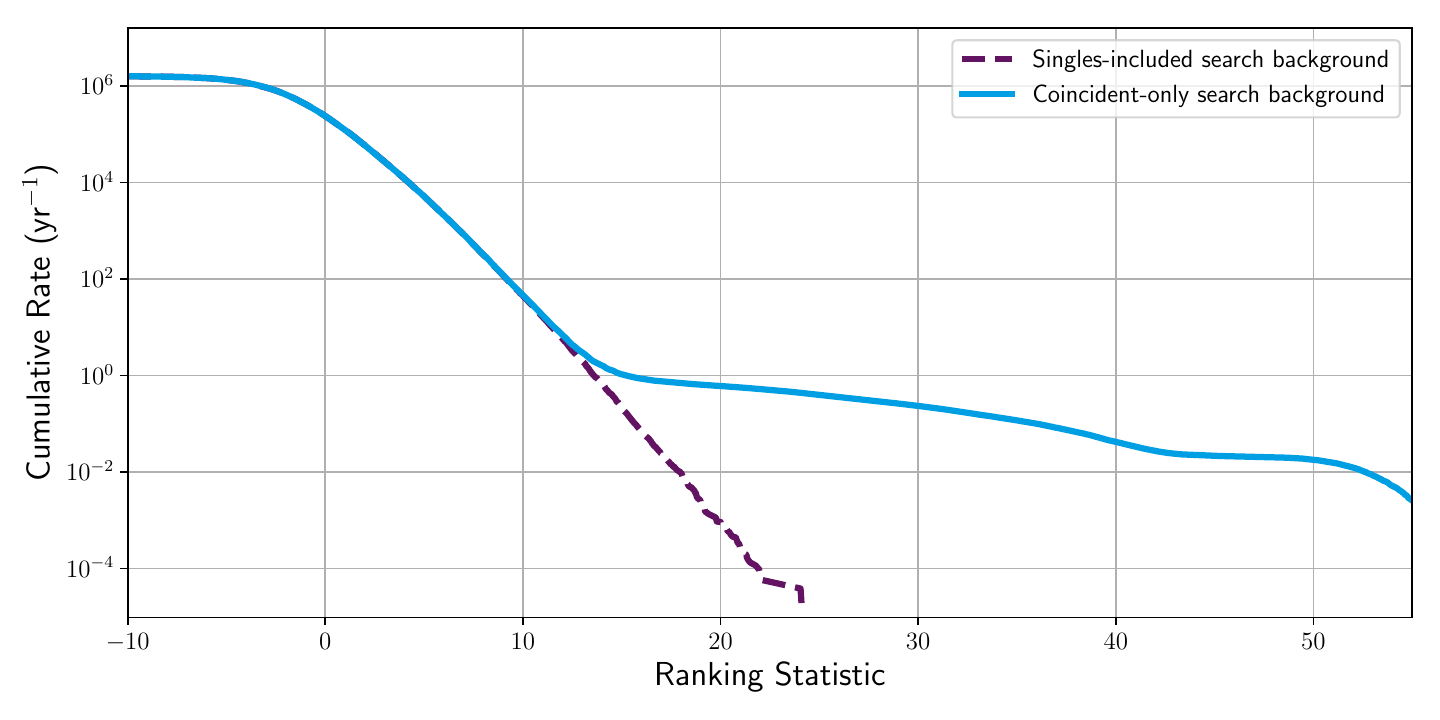}
\end{center}
\caption{\label{fig:chunk31_bg}
The \ac{LV} `exclusive' background of the analysis containing \FULLNAME{GW200105} and \FULLNAME{GW200112}, from the coincident-only search [blue, solid] and from the singles-included search where the single-detector signals are removed from the background [purple, dashed].
}
\end{figure}

Figure~\ref{fig:chunk31_bg} shows the \ac{LV} background of the analysis chunk containing \FULLNAME{GW200105} and \FULLNAME{GW200112}, where the triggers from the single-detector events in LIGO-Livingston match to random noise in Virgo, and form significant events in the background.
The result is effectively a shelf in the false alarm rate, where events cannot be seen with false alarm rates below one per a few years unless they are very significant.
We see in Figure~\ref{fig:injections_scatter_broad} that this means that events from \PYCBCBROAD are prevented from becoming more significant between \acp{FAR} of around one per 10 years and one per ten thousand years, and that this effect begins at slightly higher \ac{FAR} in \PYCBCBBH as in Figure~\ref{fig:injections_scatter_bbh}.

By injecting signals into the data, we have seen that by including single-detector events, we gain sensitivity of the search to all signals by a factor of \VTRATIOBROADALL in the \PYCBCBROAD search and \VTRATIOBBHALL in the \PYCBCBBH search, and by up to a factor of \VTRATIOBBHBINTHREE for parts of the parameter space.

\section{Conclusions}
\label{sec:conclusions}
We present a method for extrapolating the false alarm rate of gravitational wave candidate events which do not form a coincidence for use in \PYCBC searches.
This method adapts the coincident ranking statistic for use with single-detector \removed{triggers}\inserted{events}, and fits the number of \removed{triggers}\inserted{events} with that ranking statistic or higher to a falling exponential model, extrapolating the number of higher-ranked events using this exponential fit.
We have shown how this extrapolated \ac{FAR} is used in the wider context of a \ac{GW} search analysis.

We have assessed the ability of this method to digest single-detector \removed{triggers}\inserted{events} within a search.
We recover \NUMSNGLEVENTS{} single-detector events in \ac{O3} with a false alarm rate less than two per year; these events correspond to known events found by other pipelines.
Only \NUMRECOVEREDGLITCHES{} glitch was identified as a marginally significant single-detector event during \ac{O3}, showing that we have balanced the requirement to find events with the need to avoid non-Gaussian transient glitches.
The total time--volume sensitivity of the \PYCBCBROAD search increases by a factor of \VTRATIOBROADALL at a false alarm rate of one per two years compared to completely ignoring single-detector events, and the \PYCBCBBH search sensitivity increases by a factor of \VTRATIOBBHALL.

\section{Further Work}
\label{sec:further-work}

We have presented methods to extrapolate the \ac{FAR} for events in a single detector.
This method is optimal in Gaussian noise, and although we have given a description of how we can mitigate non-Gaussian glitches in the search, they can always show up as outliers, and therefore be assigned low \acp{FAR}, as evidenced by the reported significance of \FULLNAME{200218}.

Data being closer to Gaussian will then help this method even more than in a coincident search.
Ongoing efforts to understand, reduce and mitigate glitches \cite{LIGO:2020zwl,LIGO:2021ppb,Robinet:2020lbf,Soni:2021cjy,Merritt:2021xwh,Godwin:2020weu,McIsaac:2022odb}, will therefore help to improve the ability of the search to find events.

Working out the inclusive \ac{FAR} of events needs careful consideration, as we have so far only considered exclusive \ac{FAR}.
The inclusive \ac{FAR}, as noted previously, is one per live time for the loudest-ranked single-detector event.
The \ac{FAR} is added as if we have seen an event in each combination at the ranking statistic of the event - this would result in the addition of one per live time for all active detectors, and all events would then have insignificant \acp{FAR}.
The solution may be to simply state that the inclusive \ac{FAR} is one per live time of the detector it was found in; as the event we consider is, through clustering, the loudest event at that time, we do not have any higher-ranked events at that time.
However this would not follow the same process we have used up to now, and would still not be useful for finding events with useful significance.

There are also ways we can make improvements to the ranking of events in order to help us to further discriminate \removed{background}\inserted{signals} from noise.
We currently have no way to include any information from detectors which were active but did not trigger.
For example, a high-\ac{SNR} event in LIGO-Hanford is unlikely to be real if LIGO-Livingston was active but did not trigger.
Inclusion of these terms may be possible as part of an extension to the prior histograms used to calculate $\p(\vec{\Omega}|S)$, the probability of the given extrinsic parameters given that the trigger is a signal.


\section{Acknowledgements}
The authors acknowledge the \ac{STFC} for funding through grants ST/T000333/1 and ST/V005715/1.
We are grateful to Fr\'{e}d\'{e}rique Marion and Thomas Dent for help and comments on the manuscript.
This material is based upon work supported by \ac{NSF}'s LIGO Laboratory which is a major facility fully funded by the \ac{NSF}.
This research has made use of data or software available from the Gravitational Wave Open Science Center (gw-openscience.org), a service of LIGO Laboratory, the LIGO Scientific Collaboration, the Virgo Collaboration, and KAGRA. LIGO Laboratory and Advanced LIGO are funded by the United States \ac{NSF} as well as the \ac{STFC} of the United Kingdom, the Max-Planck-Society (MPS), and the State of Niedersachsen/Germany for support of the construction of Advanced LIGO and construction and operation of the GEO600 detector. Additional support for Advanced LIGO was provided by the Australian Research Council. Virgo is funded, through the European Gravitational Observatory (EGO), by the French Centre National de Recherche Scientifique (CNRS), the Italian Istituto Nazionale di Fisica Nucleare (INFN) and the Dutch Nikhef, with contributions by institutions from Belgium, Germany, Greece, Hungary, Ireland, Japan, Monaco, Poland, Portugal, Spain. The construction and operation of KAGRA are funded by Ministry of Education, Culture, Sports, Science and Technology (MEXT), and Japan Society for the Promotion of Science (JSPS), National Research Foundation (NRF) and Ministry of Science and ICT (MSIT) in Korea, Academia Sinica (AS) and the Ministry of Science and Technology (MoST) in Taiwan.
The authors are grateful for the computational resources and data provided by the LIGO Laboratory and supported by National Science Foundation Grants No. PHY-0757058 and No. PHY-0823459.
The authors also acknowledge the use of the IUCAA LDG cluster, Sarathi, for computational/numerical work.
The PyCBC offline search software used was based on PyCBC release 1.18.1~\cite{alex_nitz_2021_4849433}.
The PyCBC offline search software is built upon LALSuite~\cite{lalsuite}, numpy~\cite{numpy}, SciPy~\cite{Virtanen:2019joe}, Astropy~\cite{Price-Whelan:2018hus} and Pegasus~\cite{deelman-fgcs-2015}.

\appendix
\section{Further Results}
\label{app:further_results}
In addition to the results shown in Section~\ref{sec:o3_results}, we present in Table~\ref{tab:other_results} the \ac{FAR} and Network \ac{SNR} for events from GWTC-2.1, GWTC-3, 3-OGC and 4-OGC which were not included in Tables~\ref{tab:results_o3a} and~\ref{tab:results_o3b} .
These were not included in Tables~\ref{tab:results_o3a} or~\ref{tab:results_o3b} either as they did not meet the \ac{FAR} threshold in these analyses, or as the \ac{FAR} did not change between the analyses.

\begin{landscape}
\begingroup
\makeatletter
\rowcolors{1}{lightgray}{}
\begin{table}
\tiny
\begin{center}
\begin{tabular}{lccccccccc}
\hline
\hiderowcolors
\multicolumn{1}{c}{\multirow{4}{*}{Event}} & \multirow{2}{*}{Instruments} & \multicolumn{4}{c|}{Search including single-detector events} & \multicolumn{4}{c}{Coincident-only search} \\
 & & \multicolumn{2}{c|}{\PYCBCBROAD} & \multicolumn{2}{c|}{\PYCBCBBH} & \multicolumn{2}{c|}{\PYCBCBROAD} & \multicolumn{2}{c}{\PYCBCBBH} \\
 & \multirow{2}{*}{Active, Triggered} &  FAR   &  Network  &  FAR   & Network  &  FAR   & Network  &  FAR &  Network  \\
 & & (years$^{-1}$) & SNR & (years$^{-1}$) & SNR & (years$^{-1}$) & SNR & (years$^{-1}$) & SNR \\
\hline
\showrowcolors
{\EVENTNAMEFONT{GW190707}\FULLNAME{GW190707}} &
                \ACTIVEINSTRUMENTS{GW190707}, \ALLINSTRUMENTS{GW190707} &
                $ \SNGLBROADASSOC{GW190707} \SNGLBROADFAR{GW190707} $ &
                $ \SNGLBROADASSOC{GW190707} \SUBSETINSTRUMENTSSNGLBROAD{GW190707} \SNGLBROADNETWORKSNR{GW190707} $ &
                $ \SNGLBBHASSOC{GW190707} \SNGLBBHFAR{GW190707} $ &
                $ \SNGLBBHASSOC{GW190707} \SUBSETINSTRUMENTSSNGLBBH{GW190707} \SNGLBBHNETWORKSNR{GW190707} $ &
                $ \COINCBROADASSOC{GW190707} \COINCBROADFAR{GW190707} $ &
                $ \COINCBROADASSOC{GW190707} \SUBSETINSTRUMENTSCOINCBROAD{GW190707} \COINCBROADNETWORKSNR{GW190707} $ &
                $ \COINCBBHASSOC{GW190707} \COINCBBHFAR{GW190707} $ &
                $ \COINCBBHASSOC{GW190707} \SUBSETINSTRUMENTSCOINCBBH{GW190707} \COINCBBHNETWORKSNR{GW190707} $ \\
\end{tabular}
\end{center}
\caption{
\label{tab:other_results}
Events with \ac{FAR} unchanged between the coincident and singles-included analyses, and events with \ac{FAR} $>2$ in both the single-included and coincident searches, but which are reported in GWTC-2.1, GWTC-3, 3-OGC or 4-OGC.
The horizonatal line indicates the point where events above were recovered but unchanged, the ones below the line were not recovered in this analysis.
Events are not included if the \ac{FAR} is infinite in all searches presented here, which was the case for \EXTERNALZEROES.
Table columns and details are the same as in Table~\ref{tab:results_o3a}.
}
\end{table}
\makeatother
\endgroup

\end{landscape}

It should be noted that the event \FULLNAME{GW20021009} is indicated as an event not previously found in Table~\ref{tab:other_results}, but this was found by both \PYCBC searches in the \ac{GWTC}-3 analysis with $\pastro{} > 0.5$, the criteria for inclusion in that work.

We also provide template parameters for the events found by the singles-included search but not by the coincident search in Table~\ref{tab:template_parameters}.

\begingroup
\makeatletter
\rowcolors{1}{lightgray}{}
\begin{table}
\footnotesize
\begin{center}
\begin{tabular}{lcccccccc}
\hline
\hiderowcolors
\multicolumn{1}{c}{\multirow{2}{*}{Event}} & \multicolumn{4}{c|}{\PYCBCBROAD} & \multicolumn{4}{c}{\PYCBCBBH} \\ 
 & \mchirp & $\eta$ & $\chi_\textrm{eff}$ & Dur (s)& \mchirp & $\eta$ & $\chi_\textrm{eff}$ & Dur (s)\\
\hline
\showrowcolors
\FULLNAME{GW200112} &

{\SNGLBROADASSOC{GW200112} \SNGLBROADMCHIRP{GW200112}} &
{\SNGLBROADASSOC{GW200112} \SNGLBROADETA{GW200112}} &
{\SNGLBROADASSOC{GW200112} \SNGLBROADCHIEFF{GW200112}} &
{\SNGLBROADASSOC{GW200112} \SNGLBROADDURATION{GW200112}} &

{\SNGLBBHASSOC{GW200112} \SNGLBBHMCHIRP{GW200112}} &
{\SNGLBBHASSOC{GW200112} \SNGLBBHETA{GW200112}} &
{\SNGLBBHASSOC{GW200112} \SNGLBBHCHIEFF{GW200112}} &
{\SNGLBBHASSOC{GW200112} \SNGLBBHDURATION{GW200112}}\\
\FULLNAME{GW190425} &

{\SNGLBROADASSOC{GW190425} \SNGLBROADMCHIRP{GW190425}} &
{\SNGLBROADASSOC{GW190425} \SNGLBROADETA{GW190425}} &
{\SNGLBROADASSOC{GW190425} \SNGLBROADCHIEFF{GW190425}} &
{\SNGLBROADASSOC{GW190425} \SNGLBROADDURATION{GW190425}} &

{\SNGLBBHASSOC{GW190425} \SNGLBBHMCHIRP{GW190425}} &
{\SNGLBBHASSOC{GW190425} \SNGLBBHETA{GW190425}} &
{\SNGLBBHASSOC{GW190425} \SNGLBBHCHIEFF{GW190425}} &
{\SNGLBBHASSOC{GW190425} \SNGLBBHDURATION{GW190425}}\\
\FULLNAME{GW200105} &

{\SNGLBROADASSOC{GW200105} \SNGLBROADMCHIRP{GW200105}} &
{\SNGLBROADASSOC{GW200105} \SNGLBROADETA{GW200105}} &
{\SNGLBROADASSOC{GW200105} \SNGLBROADCHIEFF{GW200105}} &
{\SNGLBROADASSOC{GW200105} \SNGLBROADDURATION{GW200105}} &

{\SNGLBBHASSOC{GW200105} \SNGLBBHMCHIRP{GW200105}} &
{\SNGLBBHASSOC{GW200105} \SNGLBBHETA{GW200105}} &
{\SNGLBBHASSOC{GW200105} \SNGLBBHCHIEFF{GW200105}} &
{\SNGLBBHASSOC{GW200105} \SNGLBBHDURATION{GW200105}}\\
\FULLNAME{GW190708} &

{\SNGLBROADASSOC{GW190708} \SNGLBROADMCHIRP{GW190708}} &
{\SNGLBROADASSOC{GW190708} \SNGLBROADETA{GW190708}} &
{\SNGLBROADASSOC{GW190708} \SNGLBROADCHIEFF{GW190708}} &
{\SNGLBROADASSOC{GW190708} \SNGLBROADDURATION{GW190708}} &

{\SNGLBBHASSOC{GW190708} \SNGLBBHMCHIRP{GW190708}} &
{\SNGLBBHASSOC{GW190708} \SNGLBBHETA{GW190708}} &
{\SNGLBBHASSOC{GW190708} \SNGLBBHCHIEFF{GW190708}} &
{\SNGLBBHASSOC{GW190708} \SNGLBBHDURATION{GW190708}}\\
\FULLNAME{GW190620} &

{\SNGLBROADASSOC{GW190620} \SNGLBROADMCHIRP{GW190620}} &
{\SNGLBROADASSOC{GW190620} \SNGLBROADETA{GW190620}} &
{\SNGLBROADASSOC{GW190620} \SNGLBROADCHIEFF{GW190620}} &
{\SNGLBROADASSOC{GW190620} \SNGLBROADDURATION{GW190620}} &

{\SNGLBBHASSOC{GW190620} \SNGLBBHMCHIRP{GW190620}} &
{\SNGLBBHASSOC{GW190620} \SNGLBBHETA{GW190620}} &
{\SNGLBBHASSOC{GW190620} \SNGLBBHCHIEFF{GW190620}} &
{\SNGLBBHASSOC{GW190620} \SNGLBBHDURATION{GW190620}}\\
\FULLNAME{GW200302} &

{\SNGLBROADASSOC{GW200302} \SNGLBROADMCHIRP{GW200302}} &
{\SNGLBROADASSOC{GW200302} \SNGLBROADETA{GW200302}} &
{\SNGLBROADASSOC{GW200302} \SNGLBROADCHIEFF{GW200302}} &
{\SNGLBROADASSOC{GW200302} \SNGLBROADDURATION{GW200302}} &

{\SNGLBBHASSOC{GW200302} \SNGLBBHMCHIRP{GW200302}} &
{\SNGLBBHASSOC{GW200302} \SNGLBBHETA{GW200302}} &
{\SNGLBBHASSOC{GW200302} \SNGLBBHCHIEFF{GW200302}} &
{\SNGLBBHASSOC{GW200302} \SNGLBBHDURATION{GW200302}}\\
\FULLNAME{GW190910} &

{\SNGLBROADASSOC{GW190910} \SNGLBROADMCHIRP{GW190910}} &
{\SNGLBROADASSOC{GW190910} \SNGLBROADETA{GW190910}} &
{\SNGLBROADASSOC{GW190910} \SNGLBROADCHIEFF{GW190910}} &
{\SNGLBROADASSOC{GW190910} \SNGLBROADDURATION{GW190910}} &

{\SNGLBBHASSOC{GW190910} \SNGLBBHMCHIRP{GW190910}} &
{\SNGLBBHASSOC{GW190910} \SNGLBBHETA{GW190910}} &
{\SNGLBBHASSOC{GW190910} \SNGLBBHCHIEFF{GW190910}} &
{\SNGLBBHASSOC{GW190910} \SNGLBBHDURATION{GW190910}}\\
\FULLNAME{200218} &

{\SNGLBROADASSOC{200218} \SNGLBROADMCHIRP{200218}} &
{\SNGLBROADASSOC{200218} \SNGLBROADETA{200218}} &
{\SNGLBROADASSOC{200218} \SNGLBROADCHIEFF{200218}} &
{\SNGLBROADASSOC{200218} \SNGLBROADDURATION{200218}} &

{\SNGLBBHASSOC{200218} \SNGLBBHMCHIRP{200218}} &
{\SNGLBBHASSOC{200218} \SNGLBBHETA{200218}} &
{\SNGLBBHASSOC{200218} \SNGLBBHCHIEFF{200218}} &
{\SNGLBBHASSOC{200218} \SNGLBBHDURATION{200218}}\\

\hline
\hiderowcolors
\end{tabular}
\end{center}
\caption{\label{tab:template_parameters} Template parameters for events which were found in the singles-included analysis but not by the coincident-only analysis}
\end{table}
\makeatother
\endgroup

\inserted{The parameters in Table~\ref{tab:template_parameters} are not intended as parameter estimation analysis for the events; as all events here are given in GWTC-2.1 or GWTC-3, parameter estimation is given in those papers.
Parameter estimation of intrinsic parameters for events where only a single detector was operating should be largely unaffected by observation in only one detector, however extrinsic parameters such as the localisation of the source on the sky would be significantly affected~\citep{Fairhurst:2010is}.
}

\bibliographystyle{iopart-num}
\bibliography{biblio}

\end{document}